\documentclass[a4paper,11pt]{article}
\pdfoutput=1 % if your are submitting a pdflatex (i.e. if you have
\usepackage{jheppub} % for details on the use of the package, please
\usepackage[T1]{fontenc} % if needed
\usepackage{lmodern}

\usepackage{tikz}
\usetikzlibrary{positioning,decorations.pathmorphing}

\let\oldFootnote\footnote
\newcommand\nextToken\relax

\renewcommand\footnote[1]{%
    \oldFootnote{#1}\futurelet\nextToken\isFootnote}

\newcommand\isFootnote{%
    \ifx\footnote\nextToken\textsuperscript{,}\fi}

\usepackage{enumerate}

\usetikzlibrary{decorations.markings}%-------------------------------------%

\def\id{{1 \kern-.28em {\rm l}}}

\def\K3{{\bf K3}}
\def\journal#1&#2(#3){\unskip, \sl #1\ \bf #2 \rm(19#3) }
\def\andjournal#1&#2(#3){\sl #1~\bf #2 \rm (19#3) }

\def\bar{\overline}
\def\hat{\widehat}
\def\ie{{\it i.e.}}
\def\eg{{\it e.g.}}

\def\frac#1#2{{#1\over#2}}

\def\inbar{\,\vrule height1.5ex width.4pt depth0pt}
\def\IC{\relax\hbox{$\inbar\kern-.3em{\rm C}$}}
\def\IR{\relax{\rm I\kern-.18em R}}
\def\IP{\relax{\rm I\kern-.18em P}}

\catcode`\@=11
\def\slash#1{\mathord{\mathpalette\c@ncel{#1}}}
\def\underrel#1\over#2{\mathrel{\mathop{\kern\z@#1}\limits_{#2}}}

\catcode`\@=12

\def\det{{\rm det}}
\def\tr{{\rm tr}}

\def \sinh{{\rm sinh}}
\def \cosh{{\rm cosh}}

\def\det{{\rm det}}
\def\exp{{\rm exp}}

\def\ie{{\it i.e.}}
\def\eg{{\it e.g.}}

\title{ $\frac{SL(2,\mathbb{R})\times U(1)}{U(1)}$  CFT, NS5$+$F1 system and single trace $T\bar{T}$}

\author{Soumangsu Chakraborty}
\emailAdd{soumangsuchakraborty@gmail.com}

\affiliation{Department of Theoretical Physics,\\
 Tata Institute of Fundamental Research,\\
$1^{st}$ Homi Bhabha Road, Mumbai 400005, India}

\abstract{In this paper we prove the equivalence among (i) the weakly coupled worldsheet string theory described by the coset sigma model $\frac{SL(2,\mathbb{R})_k\times U(1)}{U(1)}\times S^3 \times T^4$ with $SL(2,\mathbb{R})$ WZW level $k\geq 2$, (ii) the full near horizon theory of the NS5 branes with $k$ NS5 branes wrapping $T^4\times S^1$, $p\gg1$ F1 strings wrapping $S^1$ and $n$ units of momentum along the $S^1$ and (iii) the single trace $T\bar{T}$ deformation of string theory in $AdS_3\times S^3\times T^4$.  As a check we compute the spectrum of the spacetime theory by performing BRST quantization of the coset description of the worldsheet theory and show that it matches exactly with the one derived in the case of single trace $T\bar{T}$ deformed string theory in $AdS_3$. Secondly, we compute the two-point correlation function of local operators of the spacetime theory using the worldsheet coset approach and reproduce the same two-point function from the supergravity approach.}

\begin{document}
\maketitle
\flushbottom

\section{Introduction}

With the advent of $T\bar{T}$ deformation of CFT$_2$ \cite{Smirnov:2016lqw,Cavaglia:2016oda}, there has been a considerable amount of progress being made in understanding non-AdS holography \cite{Giveon:2017nie,Giveon:2017myj,Asrat:2017tzd,Giribet:2017imm,Chakraborty:2018kpr,Chakraborty:2018aji,Apolo:2019zai,Chakraborty:2020swe,Chakraborty:2020fpt} (also see \cite{Apolo:2018qpq,Chakraborty:2018vja,Araujo:2018rho,Chakraborty:2019mdf,Apolo:2019yfj,Chakraborty:2020cgo,Chakraborty:2020udr} for further generalizations that include $J\bar{T}$ and $T\bar{J}$). In the presence of pure NS-NS H-flux with all the R-R fluxes switched off, the string theory models studied in \cite{Giveon:2017nie,Chakraborty:2020swe}, are closely related to double trace $T\bar{T}$ deformation of a CFT$_2$. Such a deformation of string theory in $AdS_3$ is often referred to as single trace $T\bar{T}$ deformation. The dual holographic background interpolates between $AdS_3$ in the IR to flat spacetime with a linear dilaton in the UV.  At finite temperature however, the background geometry interpolates between  a BTZ black hole in the IR, that is described by a particular orbifold sigma model on $\frac{SL(2,\mathbb{R})}{\Gamma}$ \cite{Carlip:2005zn} where $\Gamma$ is a discrete subgroup of $SL(2,\mathbb{R})$  and a black hole in linear dilaton spacetime in the UV described by the sigma model $\frac{SL(2,\mathbb{R})}{U(1)}\times U(1)$ \cite{Dijkgraaf:1991ba}.  The spacetime theory interpolates between a CFT$_2$ in the IR to a certain Little String Theory (LST) in the UV. The spacetime theory,  is non-local in the sense that the short distance physics is not governed by a fixed point.

As an example, the above interpolating background can be constructed as follows. Let us start with $k$ NS5 branes wrapping say $T^4\times S^1$ and then take the near horizon limit of the fivebranes. The background geometry is that of flat spacetime with a linear dilaton. The geometry contains a linear dilaton throat parametrized by $\phi$ such that the boundary is at $\phi\to\infty$. The string coupling goes to zero near the boundary whereas it diverges as $\phi\to-\infty$. Studying string dynamics away from the fivebranes requires non-perturbative physics. Next, we put into this background $p$ F1 strings wrapped along the $S^1$ with $n$ units of momentum along the $S^1$.  The addition of the F1 strings modifies the geometry in the IR (\ie\ the near horizon geometry of the F1 strings) to $AdS_3$. The string coupling stabilizes and saturates to $g_s^2\sim1/p$ close to the F1 strings. Thus for large $p$ the string coupling is small and one can trust string perturbation theory. The resulting geometry interpolates between $AdS_3$ in the IR which corresponds to the near horizon limit of the F1 strings in the setup stated above, to flat spacetime with a liner dilaton in the UV which corresponds to the near horizon geometry of just the NS5 branes (far away from the F1 strings).

A natural question that one may raise at this point is: does there exist a single solvable worldsheet description that will capture the full interpolating theory between BTZ black hole in the IR to a black hole in a linear dilaton geometry in the UV?
This question has been partially answered in the context of single trace $T\bar{T}$ deformation of string theory in $AdS_3$ at zero temperature where the background interpolates between $AdS_3$ in the IR to linear dilaton spacetime in the UV.
 In this paper we argue that the  gauged WZW model $\frac{SL(2,\mathbb{R})_k\times U(1)}{U(1)}$ with the gauge currents given by \eqref{gaugecurr} indeed describes the full interpolating bulk geometry both at zero and finite temperature.  To be more precise, we argue in this paper that the worldsheet coset CFT $\frac{SL(2,\mathbb{R})_k\times U(1)}{U(1)}\times S^3\times T^4$ with gauge currents given by \eqref{gaugecurr} and with $SL(2,\mathbb{R})$ at level $k$, the full near horizon geometry of the $k$ NS5 branes in a system of $k$ NS5 branes wrapping $T^4\times S^1$, $p$ F1 strings wrapping $S^1$ with $n$ units of momentum along the $S^1$ and the single trace $T\bar{T}$ deformation of string theory in $AdS_3\times S^3\times T^4$ both at zero and finite temperature are equivalent descriptions of each other. Evidences for the above equivalence have already appeared in many papers over last the two decades (see \eg\ \cite{Giveon:2003ge,Giveon:2005mi,Giveon:2006pr,Giveon:2019fgr,Apolo:2019zai,Chakraborty:2020swe} and references there in). In this paper we tie all the loose ends together with detailed computation and establish the above mentioned equivalence. 
 
  As a check to the above proposed equivalence, we compute the spectrum of the spacetime theory using the worldsheet coset sigma model and find that it matches exactly with the spectrum computed in the case of single trace $T\bar{T}$ deformed string theory in $AdS_3$.  Secondly, we compute the correlation of two operators (in momentum space) of the spacetime theory using the coset worldsheet approach. This matches exactly with the two-point function computed in the case of single trace $T\bar{T}$ deformed string theory in $AdS_3$ \cite{Asrat:2017tzd}. Finally, we reproduced the same correlation function from a supergravity computation. 

The coset construction of the worldsheet description of the full near horizon geometry of the NS5 branes in a system of NS5+F1+momentum (in the configuration stated above) has the following advantages. In this construction, the  symmetries of the spacetime LST is manifest. Secondly, the coset description enables us to systematically construct physical worldsheet vertex operators which otherwise may not be straight forward.  Thirdly, the BRST quantization enables us to construct the physical Hilbert space just by constructing the BRST cohomology of the gauged WZW model using standard techniques.   

Single trace $T\bar{T}$ deformation of pure NS-NS $AdS_3\times S^3\times T^4$ has also been studied from the integrability approach. Worldsheet S-matrix has been shown to reproduce the same CDD factor obtained in the case of double trace $T\bar{T}$ deformed CFT$_2$ \cite{Baggio:2018gct,Dei:2018mfl}.

This paper is organized as follows. In section \ref{sec2}, we give a brief review of $SL(2,\mathbb{R})$ current algebra, that we have used heavily in the paper. We also discuss about the theory on the long strings and the $T\bar{T}$ deformation of the theory living on a single long string.  In section \ref{sec3}, we construct the sigma model background of the coset $\frac{SL(2,\mathbb{R})_k\times U(1)}{U(1)}$ and show that it describes the full near horizon theory of the NS5 branes in a system of NS5 branes, F1 strings and momentum modes along the F1 direction.  In section \ref{sec4}, we perform a BRST quantization of the gauged WZW model under consideration, construct physical vertex operators and eventually calculate the spectrum of the spacetime theory. In section \ref{sec5}, we compute two-point correlation function of local operators of the spacetime theory from the worldsheet approach and compare with the one computed from the supergravity approach. In section \ref{sec6}, we discuss our results and list few avenues to future research.

\section{A brief review of string theory in $AdS_3$ and its single trace $T\bar{T}$ deformation}\label{sec2}

In this section, we will briefly review certain aspects of superstrings in $AdS_3$ and its single trace $T\bar{T}$ deformation which we will use heavily later in the sections that follow.  For more detailed discussion of string theory in $AdS_3$, we will refer to the reader the following references \cite{Giveon:1998ns,Kutasov:1999xu,Giveon:2001up,Maldacena:2000hw,Maldacena:2000kv,Maldacena:2001km}.

\subsection{Current algebra}

String theory in $AdS_3$ with the NS-NS H-flux turned on and all the R-R fluxes switched off is described by the WZW model on $SL(2,\mathbb{R})$ group manifold. The sigma model has an affine $SL(2,\mathbb{R})_L\times SL(2,\mathbb{R})_R$  symmetry at level $k$. The level of the $\mathfrak{sl}(2,\mathbb{R})_{L,R}$ current algebra is related to the radius of $AdS_3$ as $R_{ads}=\sqrt{k}l_s$ where $l_s=\sqrt{\alpha'}$ is the string length. For an element $g\in SL(2,\mathbb{R})_k$, the WZW sigma model action is given by
\begin{eqnarray}\label{wzwsl22r}
S=\frac{k}{4\pi}\left[\int_\Sigma d^2 z \tr\left(g^{-1}\partial g g^{-1}\bar{\partial }g\right)-\frac{1}{3}\int_B d^3X \tr\left((g^{-1}dg)^3\right) \right],
\end{eqnarray}
where $\Sigma$ is the compact worldsheet  Riemann surface, and $B$ is the three-dimensional extension of $\Sigma$ such that its boundary, $\partial B$, is $\Sigma$ (\ie\ $\partial B=\Sigma$) and $z,\bar{z}$ are the complex coordinates on $\Sigma$. The global $SL(2,\mathbb{R})_L\times SL(2,\mathbb{R})_R$ symmetries are generated by  the currents $J^A(z),\bar{J}^A(\bar{z})$ for $A=\{1,2,3\}$ given by
\begin{eqnarray}
\begin{split}
&J^A=-\eta^{AB}\frac{k}{2}\tr\left(\partial gg^{-1}\sigma_B\right),\\
& \bar{J}^A=\eta^{AB}\frac{k}{2}\tr\left(g^{-1}\bar{\partial} g\sigma_B\right),
\end{split}
\end{eqnarray}
where $\sigma_A$ are the Pauli matrices given by
\begin{eqnarray}\label{pauli}
\sigma_1=\begin{pmatrix}
0 & 1\\1 &0
\end{pmatrix}, \ \ \ \ \sigma_2=\begin{pmatrix}
0 &-i\\ i& 0
\end{pmatrix}, \ \ \ \ \sigma_3=\begin{pmatrix}
1 &0 \\ 0 & -1
\end{pmatrix}.
\end{eqnarray}
 The indices on the currents $J^A$ and $\bar{J}^A$ are raised and lowered by the metric $\eta_{AB}=\rm{diag}\{1,1,-1\}$  for Lorentzian $SL(2,\mathbb{R})$ and $\eta_{AB}=\delta_{AB}$ for Euclidean $SL(2,\mathbb{R})$.
The currents $J^A$ satisfy the following OPE algebra:
\begin{eqnarray}\label{sl2rjjope}
J^A(z)J^B(w)=\frac{\frac{k}{2}\eta^{AB}}{(z-w)^2}+\frac{if^{ABC}}{z-w}J^C(w)~,
\end{eqnarray}
where $f^{ABC}$ are the $\mathfrak{sl}(2,\mathbb{R})$ structure constants. It is often useful to define the light cone currents, $J^\pm,\bar{J}^\pm$ (in Lorentzian $AdS_3$), as
\begin{eqnarray}
J^+=J^1+iJ^2~, \ \ \ \ J^-=J^1-iJ^2~.
\end{eqnarray}
Changing the chiralities one can similarly define $\bar{J}^{\pm}$.

The holomorphic and anti-holomorphic components of the stress tensor can be constructed using Sugawara construction:
\begin{eqnarray}
\begin{split}
T(z)&=\frac{1}{k}\eta_{AB}J^AJ^B=\frac{1}{k}\left(-(J^3)^2+J^+J^-\right)~,\\
\bar{T}(\bar{z})&=\frac{1}{k}\eta_{AB}\bar{J}^A\bar{J}^B=\frac{1}{k}\left(-(\bar{J}^3)^2+\bar{J}^+\bar{J}^-\right)~.
\end{split}
\end{eqnarray}

\subsection{Long strings}

Thanks to the gauge/gravity correspondence, string theory in $AdS_3$ is dual to a CFT$_2$ living on the boundary of $AdS_3$. In the presence of pure NS-NS H-flux, the spacetime theory has a normalizable $SL(2,\mathbb{C})$ invariant vacuum: (a) the NS vacuum that corresponds to global $AdS_3$ in the bulk and (b) the R vacuum that correspond to massless BTZ black hole in the bulk. The NS sector contains states whose spectrum is discrete. These states belongs to the principal discrete series representation of $SL(2,\mathbb{R})_k$. The NS sector also contains a continuum above a gap of order $k/2$ \cite{Maldacena:2000hw}. This continuum of states belong to the continuous series representation of $SL(2,\mathbb{R})_k$. On the other hand, the R sector contains a continuum of long strings above a gap of the order of $1/k$ \footnote{To the best of our knowledge, the status of the discrete series states in the the R-sector is not clear.}. Thus for large $k$ (\ie\ in the supergravity approximation) the gap is very small. In the rest of the discussion in this paper, we will mostly focus on large $k$ such that the semiclassical bulk description holds and concentrate only on the long strings in the R sector. Of course, understanding the discussion that follows in the rest of the paper, in the NS sector is an interesting problem on its own merit. To the best of our knowledge, this issue has not yet been addressed.

The theory on a single long string was first analyzed in \cite{Seiberg:1999xz}. The authors of \cite{Seiberg:1999xz} showed that for string theory on $AdS_3\times \mathcal{N}_7$, where $\mathcal{N}_7$ is a seven-dimensional compact manifold, the theory living on a single long string is described by a sigma model on 
\begin{eqnarray}
\mathcal{M}_{6k}^{(L)}=\mathbb{R}_\phi\times \mathcal{N}_7~,
\end{eqnarray}
with central charge $c_{\mathcal{M}}=6k$, where the theory on $\mathbb{R}_\phi$ has a dilaton that is linear in the radial direction $\phi$ with slope given by
\begin{eqnarray}
Q^{(L)}=(k-1)\sqrt{\frac{2}{k}}~.
\end{eqnarray}
The effective coupling on the long strings is given by $g^{(L)}\sim \exp\left({Q^{(L)}\phi}\right)$. Thus, as the long strings move towards the boundary, their dynamics become strongly coupled \footnote{Remember that we are assuming that $k\geq 2$. For $k<1$, the coupling of the theory on the long strings is quite the opposite. They become weakly coupled as they move towards the boundary \cite{Giveon:2005mi}. At $k=1$, the story becomes more intriguing \cite{Giveon:2005mi,Eberhardt:2018ouy}.}. But there is a wide range of positions in the radial direction where the long strings are weakly coupled and one can trust string perturbation theory. The full boundary theory dual to string theory in $AdS_3$ is not known for generic $k$ but the theory on the long strings are describes by the symmetric orbifold \cite{Chakraborty:2019mdf} \footnote{It is however important to stress the fact that $\left(\mathcal{M}_{6k}^{(L)}\right)^p/S_p$ is not the full spacetime theory. For comments on possible symmetric orbifold structure of the full theory, see \cite{Chakraborty:2019mdf}.}
\begin{eqnarray}
\frac{\left(\mathcal{M}_{6k}^{(L)}\right)^p}{S_p}~,
\end{eqnarray}
where $p$ can be thought of at the number of F1 strings that forms the background.

\subsection{Single trace $T\bar{T}$ deformation}

String theory in $AdS_3$ is dual to a CFT$_2$ living on the boundary of $AdS_3$. The $T\bar{T}$ deformation is a double trace deformation of the theory. But, unfortunately string theory doesn't provide a good understanding of such a deformation. It not yet clear from the bulk side of the duality, why such a deformation is solvable. But, string theory in $AdS_3$ with NS-NS H-flux turned on and all the R-R fluxes switched off, contains an operator $D(x,\bar{x})$ \cite{Kutasov:1999xu} (where $x$ and $\bar{x}$ are the complex coordinates of the boundary theory), that has many properties in common with the $T\bar{T}$ operator. The operator $D(x,\bar{x})$ is an operator in the long string sector of the spacetime theory and is a $(2,2)$ quasi-primary operator of the spacetime Virasoro algebra and it has the same OPE with the stress tensor at the actual $T\bar{T}$ operator. But unlike the  $T\bar{T}$ operator, $D(x,\bar{x})$ is a single trace operator. However the operator $D(x,\bar{x})$ is related to the $T\bar{T}$ operator in an interesting way \cite{Kutasov:1999xu}:
\begin{eqnarray}\label{D(x)}
D(x,\bar{x})=\sum_{i=1}^pT_i(x)\bar{T}_i(\bar{x})~,
\end{eqnarray}
where $T_i\bar{T}_i$ is the $T\bar{T}$ operator of the individual  block block $\mathcal{M}^{(L)}_{6k}$ in the symmetric product $(\mathcal{M}^{(L)}_{6k})^p/S_p$. Deformation of the spacetime theory by the operator $D(x,\bar{x})$ would imply deforming the individual block $\mathcal{M}^{(L)}_{6k}$ by its own $T\bar{T}$ operator and then symmetrized. The interesting fact about the deformation of the spacetime theory by the operator $D(x,\bar{x})$ is that it induces on the dual worldsheet theory a truly marginal current-anti-current deformation \cite{Giveon:2017nie}:
\begin{eqnarray}\label{djjbar}
\int_{(\mathcal{M}^{(L)}_{6k})^p/S_p} d^2x ~D(x,\bar{x})\sim \int_\Sigma d^2x~ J^-\bar{J}^-~.
\end{eqnarray}
The deformation \eqref{djjbar} of the string theory in $AdS_3$ is often referred to as the single trace $T\bar{T}$ deformation in the literature. 
Such a worldsheet deformation is exactly solvable. The deformed worldsheet theory is given by \cite{Forste:1994wp,Israel:2003ry}
\begin{eqnarray} \label{m3}
S=\frac{k}{2\pi}\int_\Sigma d^2z \left(\partial\phi\bar{\partial}\phi+ \frac{\partial\bar{\gamma}\bar{\partial}\gamma}{\lambda+e^{-2\phi}}\right)~,
\end{eqnarray}
where $\phi$ is the radial direction, $\gamma\ \&\ \bar{\gamma}$ are the transverse light cone directions and $\lambda$ is the truly marginal worldsheet coupling \footnote{By appropriate scaling and shift of the coordinates, one can set  $\lambda=1$  \cite{Giveon:2017nie}.}. In the discussion that follows, we will consider $\lambda>0$. For a detailed discussion on the deformation by negative $\lambda$, see \cite{Chakraborty:2020swe}. The above sigma model \eqref{m3} interpolates between $AdS_3$ in the IR to flat spacetime with a linear dilaton in the UV.  For further details on the various observables in this background see \cite{Giribet:2017imm,Asrat:2017tzd,Chakraborty:2018kpr,Chakraborty:2018aji}.

As an example of the above interpolating geometry one may consider a stack of $k$ NS5 branes wrapping $T^4\times S^1$ and $p$ F1 strings wrapping the $S^1$. Going to the near horizon geometry of the NS5 branes one obtains flat spacetime with a linear dilaton. Then approaching the near horizon geometry of the F1 strings, one obtains $AdS_3$. The full interpolating background \eqref{m3} (interpolating between $AdS_3$ in the IR to linear dilaton spacetime in the UV) corresponds to interpolation between the near horizon geometry of the F1 strings in the IR to the near horizon geometry of just the NS5 branes in the UV. In section \ref{sec3} we will develop a coset description that  describes the full near horizon theory of the NS5 branes in a system of NS5 branes plus F1 strings with certain units of momentum modes along the F1 direction at any temperature.

\section{$\frac{SL(2,\mathbb{R})_k\times U(1)}{U(1)}$ coset CFT}\label{sec3}

In this section, we will derive the $\frac{SL(2,\mathbb{R})_k\times U(1)}{U(1)}$ sigma model action and show that this sigma model background is exactly same as the full near horizon background of a stack of $k$ NS5 branes in a system of $k$ NS5 branes wrapping $T^4\times S^1$ and $p$ F1 strings wrapping $S^1$ with $n$ units of momentum along the $S^1$ at any temperature. This $\frac{SL(2,\mathbb{R})_k\times U(1)}{U(1)}$ coset sigma model had also been studied in \cite{Giveon:2003ge} in the context of scattering from beyond the singularity of a two-dimensional black hole, in \cite{Goykhman:2013oja} in the context of analyzing the hydrodynamics of some $1+1$-dimensional quantum systems and in \cite{Apolo:2019zai} in the context of single trace $T\bar{T}$ deformation and and its relation to T-duality-shift-T-duality (TsT).\footnote{We thank Luis Apolo for pointing out reference \cite{Apolo:2019zai} that studies $\frac{SL(2,\mathbb{R})_k\times U(1)}{U(1)}$ sigma model description in the context of single trace $T\bar{T}$ deformation and its relation to TsT and current-anti-current deformation of the worldsheet theory.} It has been shown in \cite{Apolo:2019zai} that the coset construction exactly reproduces the single trace $T\bar{T}$ deformed background.

\subsection{$\frac{SL(2,\mathbb{R})_k\times U(1)}{U(1)}$ action}

Let $(g,x)\in SL(2,\mathbb{R})\times U(1)$ be a point on the group manifold $SL(2,\mathbb{R})\times U(1)$. Let the $U(1)$ in $SL(2,\mathbb{R})\times U(1)$ be compact and parametrized by the coordinate $x$ such that $x\sim x+2\pi R$. The $U(1)$ scalar field $x$ is normalized such that 
\begin{eqnarray}
x(z)x(w)=-\frac{1}{2}\log (z-w)~.
\end{eqnarray}
The $U(1)_{L/R}$ currents corresponding to shifts in $x_{L,R}$,
\begin{eqnarray}
J_{x}=i\partial x, \ \ \ \ \bar{J}_x=i\bar{\partial} x ~,
\end{eqnarray}
satisfies the following OPE algebra
\begin{eqnarray}\label{u1ope}
J_x(z)J_x(w)=\frac{1/2}{(z-w)^2}~.
\end{eqnarray}

 We would like to gauge the $U(1)$ subgroup of $SL(2,\mathbb{R})\times U(1)$ that acts on it as
\begin{eqnarray}\label{gauget}
(g,x_L,x_R)\sim\left(e^{\frac{1}{\sqrt{k}}\rho_1\sigma_3}ge^{\frac{1}{\sqrt{k}}\tau_1\sigma_3},x_L+\rho_2,x_R+\tau_2\right)~. 
\end{eqnarray}
where holomorphic and anti-holomorphic decomposition of $x$ is given by $x(z,\bar{z})=x_L(z)+x_R(\bar{z})$. 
Since we are gauging only a $U(1)_L\times U(1)_R$ subgroup of $U(1)_L\times U(1)_R\times U(1)_L\times U(1)_R$ (action of $U(1)_L\times U(1)_R\times U(1)_L\times U(1)_R$ is given by \eqref{gauget}) , the parameters $\tau_{1,2}$ are not independent and $\rho_{1,2}$ can be expressed as a linear combination of $\tau_{1,2}$. Since $\tau_{1,2}$ are not independent, one can write 
\begin{eqnarray}\label{tauu}
\vec{\tau}=\begin{pmatrix}
\tau_1\\ \tau_2
\end{pmatrix}=\tau \vec{u}~,
\end{eqnarray} 
where $\tau=|\vec{\tau}|=\sqrt{\tau_1^2+\tau_2^2}\in \mathbb{R}$  and $\vec{u}$ is a constant vector. Without loss of generality, one can choose 
\begin{eqnarray}\label{u}
\vec{u}=\begin{pmatrix}
\cos\chi \\ \sin\chi
\end{pmatrix}.
\end{eqnarray}

Let us represent an element  $G\in SL(2,\mathbb{R})\times U(1)$ as 
\begin{eqnarray}\label{G}
G=\begin{pmatrix}
g & 0\\
0 & e^{\sqrt{\frac{2}{k}}x}
\end{pmatrix}.
\end{eqnarray}
The WZW action for $G\in SL(2,\mathbb{R})\times U(1) $ is given by
\begin{eqnarray}\label{wzw1}
S[G]=\frac{k}{4\pi}\left[\int_\Sigma d^2 z \tr\left(G^{-1}\partial G G^{-1}\bar{\partial }G\right)-\frac{1}{3}\int_B d^3X \tr\left((G^{-1}dG)^3\right) \right],
\end{eqnarray}
where $\Sigma$ is the compact worldsheet  Riemann surface, and $B$ is the three-dimensional extension of $\Sigma$ such that $\partial B=\Sigma$. In terms of the fields $g$ and $x$, the action \eqref{wzw1} takes the following form
\begin{eqnarray}\label{wzw2}
S=\frac{k}{4\pi}\left[\int_\Sigma d^2 z \tr\left(g^{-1}\partial g g^{-1}\bar{\partial }g\right)-\frac{1}{3}\int_B d^3X \tr\left((g^{-1}dg)^3\right) \right]+\frac{1}{2\pi}\int_\Sigma d^2z \partial x\bar{\partial} x~.
\end{eqnarray}

The gauge transformation \eqref{gauget} acts on $G$ as
\begin{eqnarray}\label{Gt}
G\to e^{T_L}Ge^{T_R},
\end{eqnarray}
where $T_{L,R}$ are respectively given by
\begin{eqnarray}\label{tltr}
T_L=\begin{pmatrix}
e^{\frac{1}{\sqrt{k}}\rho_1\sigma_3} & 0\\
0 & e^{\frac{1}{\sqrt{k}}\rho_2}
\end{pmatrix}, \ \ \ \ \ T_R=\begin{pmatrix}
e^{\frac{1}{\sqrt{k}}\tau_1\sigma_3} & 0\\
0 & e^{\frac{1}{\sqrt{k}}\tau_2}
\end{pmatrix}~.
\end{eqnarray}
For an anomaly free gauging one requires
\begin{eqnarray}\label{anomaly}
\tr(T_L^2)=\tr (T_R^2)~.
\end{eqnarray}
This would impose the condition that $\rho\equiv \sqrt{\rho_1^2+\rho_2^2}=|\vec{\rho}|=|\vec{\tau}|=\tau$. Thus without loss of generality one can write 
\begin{eqnarray}\label{rhotau}
\vec{\rho}=R\vec{\tau}~,
\end{eqnarray}
where $R$ is an $SO(2)$ matrix parametrized by
\begin{eqnarray}\label{Rmatrix}
R=\begin{pmatrix}
\cos\psi & \sin\psi\\
-\sin\psi & \cos\psi
\end{pmatrix}~.
\end{eqnarray}

Next, to perform the gauging \eqref{gauget} (with $\vec{\tau}$ and $\vec{\rho}$ related to each other via \eqref{tauu} and \eqref{rhotau}), one lifts the parameters $\vec{\tau},\vec{\rho}$ to dynamical fields $\vec{\hat{\tau}}(z,\bar{z}),\vec{\hat{\rho}}(z,\bar{z})$ subject to the constraints
\begin{eqnarray}
\vec{\hat{\tau}}(z,\bar{z})=\hat{\tau}(z,\bar{z}) \vec{u}~, \ \ \ \ \ \vec{\hat{\rho}}(z,\bar{z})=\hat{\tau}(z,\bar{z}) R\vec{u}~.
\end{eqnarray}
Thus the gauged WZW action is given by \cite{Giveon:2003ge}
\begin{eqnarray}\label{gwzw1}
\begin{split}
S_g[g,x,\hat{\tau}]&=S\left[e^{\frac{1}{\sqrt{k}}\hat{\rho}_1\sigma_3}ge^{\frac{1}{\sqrt{k}}\hat{\tau}_1\sigma_3}\right]+ S[x_L+\hat{\rho}_2,x_R+\hat{\tau}_2]\\
& \ \ \ \ \ \ -\frac{1}{2\pi} \int_\Sigma d^2 z \left(\partial \vec{\hat{\rho}}-R\partial \vec{\hat{\tau}}\right)^T\left(\bar{\partial} \vec{\hat{\rho}}-R\bar{\partial} \vec{\hat{\tau}}\right)~.
\end{split}
\end{eqnarray}
The gauged WZW action \eqref{gwzw1} is invariant under the gauge transformation $\eqref{gauget}$ with \eqref{tauu} and \eqref{rhotau} along with the following transformation of the fields
\begin{eqnarray}\label{gauget1}
\vec{\hat{\tau}}(z,\bar{z}) \to \vec{\hat{\tau}}(z,\bar{z})-\vec{\tau}~, \ \ \ \ \  \vec{\hat{\rho}}(z,\bar{z}) \to \vec{\hat{\rho}}(z,\bar{z})-\vec{\rho}~.
\end{eqnarray}
Using the Polyakov-Wiegmann identity
\begin{eqnarray}\begin{split}\label{PW}
S[UGV]=&S[G]+S[U]+S[V]\\
&+\frac{k}{2\pi}\int_\Sigma d^2z \tr\left(G^{-1}\bar{\partial}G\partial V V^{-1}+U^{-1}\bar{\partial}U\partial G G^{-1}+U^{-1}\bar{\partial}U G\partial V V^{-1}G^{-1}\right),
\end{split}
\end{eqnarray}
the gauged WZW action \eqref{gwzw1} can be written as \cite{Karabali:1989dk}
\begin{eqnarray}\label{gwzw2}
S_g[g,x,A,\bar{A}]=S[g]+S[x]+\frac{1}{2\pi}\int_\Sigma d^2 z\left(A\bar{\bf{J}}+\bar{A}{\bf{J}}+2A\bar{A}(R\vec{u})^TM\vec{u}\right)~,
\end{eqnarray}
where 
\begin{eqnarray}
\begin{split}\label{sgsx}
&S[g]=\frac{k}{4\pi}\left[\int_\Sigma d^2 z~ \tr\left(g^{-1}\partial g g^{-1}\bar{\partial }g\right)-\frac{1}{3}\int_B d^3X~ \tr\left((g^{-1}dg)^3\right) \right]~,\\
&S[x]= \frac{1}{2\pi}\int_\Sigma d^2z \partial x\bar{\partial} x~.
\end{split}
\end{eqnarray}
The gauge fields $A,\bar{A}$ are given by
\begin{eqnarray}
\begin{split}\label{AAb}
&A=i\vec{u}^T\partial \vec{\hat{\tau}}=i\partial \hat{\tau}_L~,\\
& \bar{A}=-i (R\vec{u})^T\bar{\partial}\vec{\hat{\rho}}=-i\bar{\partial}\hat{\tau}_R~,
\end{split}
\end{eqnarray}
where $\hat{\tau}(z,\bar{z})=\hat{\tau}_L(z)+\hat{\tau}_R(\bar{z})$
and the $U(1)$ gauge currents are given by
\begin{eqnarray}\label{gaugecurr}
\begin{split}
&{\bf J}=i\left(\sqrt{k}~\tr (\partial gg^{-1}\sigma_3),2\partial x\right)(R\vec{u})=\left(\frac{2i}{\sqrt{k}}J^3,2J_x \right)(R\vec{u})=\frac{2i}{\sqrt{k}}\cos(\chi-\psi)J^3+2\sin(\chi-\psi)J_x ~,\\
&\bar{{\bf J}}= -i\left(\sqrt{k}~\tr (g^{-1}\bar{\partial}g \sigma_3),2\bar{\partial} x\right)\vec{u}=\left(\frac{2i}{\sqrt{k}}\bar{J}^3,-2\bar{J}_x \right)\vec{u}=\frac{2i}{\sqrt{k}}\cos\chi\bar{J}^3-2\sin\chi \bar{J}_x~,
\end{split}
\end{eqnarray}
and the $2\times 2$ matrix $M$ is given by
\begin{eqnarray}\label{matrixM}
M=\begin{pmatrix}
\frac{1}{2}\tr (g^{-1}\sigma_3 g\sigma_3) & 0\\
0 & 1
\end{pmatrix}+R~.
\end{eqnarray}
Using \eqref{sl2rjjope} and \eqref{u1ope}, one can show that the gauge currents \eqref{gaugecurr} satisfy the following OPE algebra:
\begin{eqnarray}
{\bf{J}}(z){\bf{J}}(w)=\frac{2}{(z-w)^2}~, \ \ \ \ \bar{{\bf{J}}}(\bar{z})\bar{{\bf{J}}}(\bar{w})=\frac{2}{(\bar{z}-\bar{w})^2}~.
\end{eqnarray}
The gauged action \eqref{gwzw2} is invariant under \eqref{gauget} implying that \eqref{gwzw2} is not a gauge fixed action. Thus integrating out $A,\bar{A}$ will give an action that depends only on $g,x$ and invariant under \eqref{gauget}. Upon fixing gauge we get a sigma model that has three-dimensional geometrical spacetime interpretation. Integrating out $A,\bar{A}$ one obtains 
\begin{eqnarray}\label{gwzw3}
S_g=S[g]+S[x]-\frac{1}{4\pi}\int_\Sigma d^2z \left[\frac{{\bf{J}}\bar{{\bf J}}}{(R\vec{u})^T(M\vec{u})}\right]~.
\end{eqnarray}
There is also a dilaton that takes the following form
\begin{eqnarray}\label{dilaton}
\Phi=\Phi_0-\frac{1}{2}\log\left((R\vec{u})^T(M\vec{u})\right)~,
\end{eqnarray}
where the dilaton is normalized such that $e^{\Phi_0}$ is the string coupling in flat space. In deriving \eqref{gwzw3} and \eqref{dilaton}, we have explicitly assumed that $(R\vec{u})^T(M\vec{u})\neq0$ else the whole gauging procedure will break down. Strictly speaking \eqref{gwzw3} holds at leading order in $1/k$ (\ie\ at order $\alpha'$), but it has been shown in \cite{Tseytlin:1993my} that superconformal extension of such coset backgrounds are in fact exact.   

\subsection{The sigma model background}
 
 To evaluate the action \eqref{gwzw3} one needs to specify particular parametrization of $SL(2,\mathbb{R})$. For $g\in SL(2,\mathbb{R})$, let us choose the following parametrization (see appendix \ref{appA}):
 \begin{eqnarray}\label{gpara1}
 g(\alpha,\beta,\theta)=e^{\alpha\sigma_3}g(\theta)e^{\beta\sigma_3}~,
 \end{eqnarray}
where
\begin{eqnarray}\label{gpara2}
g_\theta(\theta)=\begin{pmatrix}
\cosh\theta & \sinh\theta \\
\sinh\theta & \cosh\theta
\end{pmatrix}~.
\end{eqnarray}
It is easy to check that $\det(g)=1$. 

The gauge symmetry in the action \eqref{gwzw3} is fixed (for $\chi\neq (2n+1)\pi/2,\psi\neq (2m+1)\pi, \ \forall n,m\in \mathbb{Z}$)\footnote{For $\chi= (2n+1)\pi/2, \ \psi = (2m+1)\pi \ \forall n,m\in \mathbb{Z}$, $\Delta_\theta=(R\vec{u})^T(M\vec{u})=0$ and the whole gauging procedure breaks down.} by setting
\begin{eqnarray}\label{gfixing}
\alpha=-\beta=\frac{y}{2}~.
\end{eqnarray}
Plugging $g$ \eqref{gpara1},\eqref{gpara2} in \eqref{gwzw3}  and imposing the gauge fixing condition \eqref{gfixing} one obtains the gauge fixed action as
\begin{eqnarray}\label{gfwzw1}
S_{gf}=\frac{1}{2\pi}\int d^2z~\partial x\bar{\partial} x+\frac{k}{2\pi}\int d^2z \left(\partial \theta \bar{\partial} \theta-\sinh^2\theta \partial y \bar{\partial} y\right)-\frac{1}{4\pi}\int d^2 z ~\frac{{\bf{J}}\bar{{\bf J}}}{\Delta_\theta}~,
\end{eqnarray}
where the gauge currents \eqref{gaugecurr} assume the following forms
\begin{eqnarray}
\begin{split}\label{jjbar1}
&{\bf J}= -2i\sqrt{k}\cos(\chi-\psi)\sinh^2\theta\partial y+2i\sin(\chi-\psi)\partial x~,\\
& \bar{{\bf J}}= -2i\sqrt{k}\cos\chi\sinh^2\theta\bar{\partial} y-2i\sin\chi \bar{\partial} x~,
\end{split}
\end{eqnarray}
and 
\begin{eqnarray}\label{deltatheta}
\Delta_\theta=(R\vec{u})^T(M\vec{u})=1+\cosh^2\theta\cos\psi+\sinh^2\theta\cos(2\chi-\psi)~.
\end{eqnarray}
Rescaling $x\to\sqrt{k}x$ and plugging \eqref{jjbar1} in \eqref{gfwzw1} one obtains
\begin{eqnarray}
\begin{split}\label{gfwzw2}
S_{gf}=&\frac{k}{2\pi}\int d^2z~\Bigg{[} \partial\theta\bar{\partial}\theta+\left(1-\frac{2\sin\chi\sin(\chi-\psi)}{\Delta_\theta}\right)\partial x\bar{\partial} x-\left(\frac{2\cos(\psi/2)\sinh^2\theta}{\Delta_\theta}\right)\partial y\bar{\partial} y \\
&-\left(\frac{2\cos\chi\sin(\chi-\psi)\sinh^2\theta}{\Delta_\theta}\right)\partial x\bar{\partial} y +\left(\frac{2\sin\chi\cos(\chi-\psi)\sinh^2\theta}{\Delta_\theta}\right)\partial y\bar{\partial} x\Bigg{]}~.
\end{split}
\end{eqnarray}
Using standard worldsheet techniques, one can read off the metric, the anti-symmetric 2-form $B$ field and the dilaton as
\begin{eqnarray}\label{bkg1}
\begin{split}
\frac{ds^2}{kl_s^2}= &d\theta^2 +\left(1-\frac{2\sin\chi\sin(\chi-\psi)}{\Delta_\theta}\right)dx^2-\left(\frac{2\cos(\psi/2)\sinh^2\theta}{\Delta_\theta}\right)dy^2
+\left(\frac{2\sinh^2\theta\sin\psi}{\Delta_\theta}\right)dxdy~,\\
B=&-kl_s^2\frac{\sin(2\chi-\psi)\sinh^2\theta}{\Delta_\theta}dx \wedge dy~,\\
e^{-2\Phi}=&e^{-2\Phi_0}\Delta_\theta~.
\end{split}
\end{eqnarray}

Next, let us perform the following change of coordinates: \footnote{Note that the coordinate transformations in \eqref{coortrfn} are ill-defined at $\chi=(2n+1)\pi/2, \ \forall n\in\mathbb{Z}$:  the right hand side of the first equation in  \eqref{coortrfn} becomes independent of $\theta$, and the right hand sides of the second and third equations in  \eqref{coortrfn} blows up.}
\begin{eqnarray}\label{coortrfn}
\begin{split}
&\rho^2=kl_s^2\left(\Delta_\theta-2\sin\chi\sin(\chi-\psi)\right)~,\\
&t=\sqrt{k}l_s\frac{\cos(\psi/2)\cos(\chi-\psi/2)}{\cos\chi\cos(\chi-\psi)}y~,\\
& x\to \sqrt{k}l_sx-\frac{\sin\psi}{2\cos\chi\cos(\chi-\psi)}t~.
\end{split}
\end{eqnarray}
In these new coordinates, the sigma model background \eqref{bkg1} takes the form 
\begin{eqnarray}\label{bkg2}
\begin{split}
&ds^2=-\frac{(\rho^2-\rho_-^2)(\rho^2-\rho_-^2)}{\ell^2\rho^2}dt^2+\frac{kl_s^2\rho^2}{(\rho^2-\rho_-^2)(\rho^2-\rho_-^2)}d\rho^2+\frac{\rho^2}{\ell^2}\left(dx-\frac{\rho_+\rho_-}{\rho^2}dt\right)^2~,\\
&H= dx\wedge dt\wedge d\left(\frac{2kl_s^2\sin(\chi-\psi/2)\cos(\psi/2)}{\ell^2}\right)~,\\
& e^{2\Phi}=e^{2\Phi_0}\frac{kl_s^2}{\ell^2}~,
\end{split}
\end{eqnarray}
where $H=dB$ is the three form field strength, $\rho_{\pm}$ are the locations of the inner and outer horizons given by
\begin{eqnarray}\label{horizons}
\begin{split}
& \rho_+^2=kl_s^2\left(1+\cos\psi-2\sin\chi\sin(\chi-\psi)\right)~, \\ 
& \rho_-^2=kl_s^2\left(1-\cos(\psi)\right)~,
\end{split}
\end{eqnarray}
and $\ell$ is given by
\begin{eqnarray}\label{ell}
\ell^2=\rho^2+2kl_s^2\sin\chi\sin(\chi-\psi)~.
\end{eqnarray}
The background \eqref{bkg2} interpolates between a rotating BTZ black hole in the IR to a rotating black hole in the linear dilaton background in the UV.

\subsubsection{Certain special limits}

\noindent{\bf Zero temperature limit:} Setting $\psi=2m\pi,\chi=(2n+1)\pi/2$, $\forall m,n\in \mathbb{Z}$ in \eqref{bkg2}, the sigma model action of the coset CFT $\frac{SL(2,\mathbb{R})_k\times U(1)}{U(1)}$ is given by \eqref{m3}
\begin{eqnarray}\label{ttbarsigm}
S_g=\frac{k}{2\pi}\int_\Sigma d^2z \left(\partial\phi\bar{\partial}\phi+ \frac{\partial\bar{\gamma}\bar{\partial}\gamma}{1+e^{-2\phi}}\right)~,
\end{eqnarray}
where
\begin{eqnarray}
\phi=\ln\left(\frac{\rho}{\sqrt{2k}l_s}\right), \ \ \ \ \gamma= \frac{x+t}{\sqrt{k}l_s}, \ \ \ \ \bar{\gamma}= \frac{x-t}{\sqrt{k}l_s}~.
\end{eqnarray}
The background \eqref{ttbarsigm} \footnote{The background \eqref{ttbarsigm} is often referred to in the literature as $\mathcal{M}_3$.} is precisely what one would obtain upon deformation of the worldsheet sigma model action in $AdS_3$ by the single trace $T\bar{T}$ operator \cite{Giveon:2017nie}. Thus the coset CFT $\frac{SL(2,\mathbb{R})_k\times U(1)}{U(1)}$ can be visualized as the generalization of single trace $T\bar{T}$ deformation of the  worldsheet string theory in $AdS_3$ at finite temperature  \cite{Chakraborty:2020swe}. To provide more evidence to this statement we will show, in the next subsection,  that  the background \eqref{bkg2} is same as the background obtained from the near horizon geometry of  a stack of $k$ NS5 branes that contains $p\gg1$ F1 strings and $n$ units of momentum along the compact direction on which winds the F1 strings \cite{Chakraborty:2020swe}.

An interesting point to note is that taking the limit $\psi=2m\pi,\chi=(2n+1)\pi/2$ $\forall m,n\in \mathbb{Z}$ before and after the gauge fixing \eqref{gfixing} doesn't commute. If the limit is taken after gauge fixing, it gives the background \eqref{ttbarsigm} as discussed above. Now, let us try to understand what happens if we take the limit $\psi=2m\pi,\chi=(2n+1)\pi/2$ before gauge fixing. The gauge currents \eqref{gaugecurr} take the forms 
\begin{eqnarray}
{\bf J}=(-1)^n2J_x~,\ \ \ \ \ \  {\bf \bar{J}}=-(-1)^n2\bar{J}_x~.
\end{eqnarray}
 This means that the $U(1)$ subgroup that we want to gauge acts only on the compact $U(1)$ parametrized by $x$. In that case, intuitively we expect to get pure $SL(2,\mathbb{R})_k$ WZW model. It is easy to see that  in this limit
\begin{eqnarray}
S[x]-\frac{1}{4\pi}\int_\Sigma d^2z \left[\frac{{\bf{J}}\bar{{\bf J}}}{(R\vec{u})^T(M\vec{u})}\right]=0~.
\end{eqnarray}
 We stress the fact that this equation holds even before fixing any gauge. Thus from \eqref{gwzw3} it follows that 
 \begin{eqnarray}
 S_g=S[g]~,
 \end{eqnarray}
 implying that one recovers the pure $SL(2,\mathbb{R})_k$ WZW model as expected.

 \vspace{.2cm}
 
\noindent {\bf Extremal limit:} The extremal limit (\ie\ $\rho_+=\rho_-\equiv \rho_0$) is obtained by setting 
\begin{eqnarray}
\cos\chi\cos(\chi-\psi)=0~.
\end{eqnarray}
This would imply
\begin{eqnarray}
\begin{split}
\text{either} \ \ \ \ \ & \chi=(2n+1)\frac{\pi}{2}, \ \ \ \ \ \ \ \ \ \ \ \ \forall n\in\mathbb{Z}~,\\
\text{or} \ \ \ \ \ \ \ \ \ \ & \chi=(2n+1)\frac{\pi}{2}+\psi, \ \ \ \ \ \   \forall n\in\mathbb{Z}~.
\end{split}
\end{eqnarray}
In this limit, the background geometry takes the following form
\begin{eqnarray}\label{bkgext}
\begin{split}
&ds^2=-\frac{(\rho^2-\rho_0^2)^2}{\rho^2(\rho^2+2kl_s^2\cos\psi)}dt^2+\frac{kl_s^2\rho^2}{(\rho^2-\rho_0^2)^2}d\rho^2+\frac{\rho^2}{(\rho^2+2kl_s^2\cos\psi)}\left(dx-\frac{\rho_0^2}{\rho^2}dt\right)^2,\\
&H= dx\wedge dt\wedge d\left(\frac{2kl_s^2(-1)^n\cos^2(\psi/2)}{\ell^2}\right),\\
& e^{2\Phi}=e^{2\Phi_0}\frac{kl_s^2}{(\rho^2+2kl_s^2\cos\psi)}~,
\end{split}
\end{eqnarray}
where the horizon $\rho_0$ is given by
\begin{eqnarray}
\rho_0^2=kl_s^2(1-\cos\psi)~.
\end{eqnarray}
The background \eqref{bkgext} interpolates between an extremal BTZ in the IR to an extremal black hole in the linear dilaton background in the UV.

\subsection{NS5+F1+momentum system}\label{ses3.3}

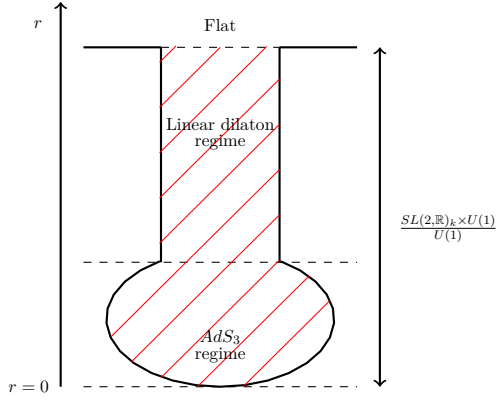
\begin{figure}
\begin{center}
\begin{tikzpicture}[scale=.6, transform shape]
  \draw [black,thick,domain=122:418] plot ({2.5*cos(\x)}, {1.5*sin(\x)});
  \draw [black,thick,domain=1.25:6] plot ({1.3}, {\x}); 
    \draw [black,thick,domain=1.25:6] plot ({-1.3}, {\x}); 
    \draw [black,dashed,-,domain=-1.3:1.3] plot ({\x}, {6}); 
     \draw [black,thick,-,domain=-3:-1.3] plot ({\x}, {6}); 
      \draw [black,thick,-,domain=1.3:3] plot ({\x}, {6}); 
    \draw [dashed,black,thin,domain=-3:3] plot ({\x}, {1.25});
    \draw [dashed,black,thin,domain=-3:3] plot ({\x}, {-1.5});
    \draw [thick,<->] (3.5,-1.5) -- (3.5,6);
     \draw [thick,->] (-3.5,-1.5) -- (-3.5,7);
      \draw [red,thin,domain=-1.95:1.3] plot ({\x}, {1+\x}); 
                \draw [red,thin,domain=-2.4:1.3] plot ({\x}, {2+\x}); 
                 \draw [red,thin,domain=-1.3:1.3] plot ({\x}, {\x}); 
                  \draw [red,thin,domain=-.5:1.95] plot ({\x}, {-1+\x}); 
                          \draw [red,thin,domain=.5:2.4] plot ({\x}, {-2+\x}); 
                          \draw [red,thin,domain=.-1.33:1.3] plot ({\x}, {3+\x}); 
                          \draw [red,thin,domain=.-1.33:1.3] plot ({\x}, {4+\x}); 
                          \draw [red,thin,domain=.-1.33:1.02] plot ({\x}, {5+\x}); 
                           \draw [red,thin,domain=.-1.33:0] plot ({\x}, {6+\x}); 
                            \draw [red,thin,domain=.-1.33:-.97] plot ({\x}, {7+\x});
      \node at (-0,3.9) {\text{regime}};
       \node at (-0,4.3) {\text{Linear dilaton}};
      \node at (0,-.4) {$AdS_3$};
      \node at (0,-.8) {\text{regime}};
           \draw (-4.2,-1.5) node {$r=0$};
           \draw (-4,6.5) node {$r$};
             \node at (-0,6.5) {\text{Flat}};
              \node at (5,2) {$\frac{SL(2,\mathbb{R})_k\times U(1)}{U(1)}$}; 
\end{tikzpicture}
\caption{The figure schematically shows the three distinct asymptotic regions of NS5+F1 system: (a) the asymptotic region far away from the NS5 branes is ten-dimensional flat space, (b) the intermediate regime namely the near horizon geometry of the NS5 branes which is the linear dilaton regime and (c) the near horizon geometry of the F1 strings where the geometry is $AdS_3$. The full near horizon theory of the NS5 branes (\ie\ the shaded region in red) is described by the gauged WZW model $\frac{SL(2,\mathbb{R})_k\times U(1)}{U(1)}$ with the gauge currents given by  \eqref{gaugecurr}.}
  \label{spacetime}
  \end{center}
\end{figure}

In this subsection, we will show that the background \eqref{bkg2} obtained form the sigma model on $\frac{SL(2,\mathbb{R})_k\times U(1)}{U(1)}$ is same as the background obtained from taking the near horizon geometry of the NS5 branes in a system of $k$ NS5 branes and $p$ F1 strings with $n$ units of momentum along the compact direction on which wraps the F1 strings at any temperature. Pictorially this corresponds to the region shaded in red, in figure \ref{spacetime}.

Let us consider type II superstrings on $\mathbb{R}^{1,4}\times S^1\times T^4$ that contains $k$ NS5 branes wrapping $T^4\times S^1$ and $p$ F1 strings winding along the $S^1$ and $n$ units of momentum along the $S^1$. The metric, the antisymmetric 3-form field strength $H$ and the dilaton are given by \cite{Chakraborty:2020swe}
\begin{eqnarray}\label{ns5f1}
\begin{split}
&ds^2=\frac{1}{f_1}\left[-\frac{f}{f_n}dt^2+f_n\left(dx-\frac{r_0^2\sinh2\alpha_n}{2f_nr^2}dt\right)^2\right]+f_5\left(\frac{1}{f}dr^2+r^2d\Omega_3^2\right)+ds^2_{T^4}~,\\
&H=dx\wedge dt\wedge d\left(\frac{r_0^2\sinh2\alpha_1}{2f_1r^2}\right)+r_0^2\sinh2\alpha_5d\Omega_3~,\\
& e^{2\Phi}=g^2\frac{f_5}{f_1}~,
\end{split}
\end{eqnarray}
where $g^2=e^{2\Phi(r\to\infty)}$ is the asymptotic string coupling. The circle $S^1$ is parametrized by the coordinate $x$ such that $x\sim x+2\pi R$, $\Omega_3$ are the spherical coordinates of the three sphere that surrounds the fivebranes,  $r$ is the radius of this three sphere and $ds^2_{T^4}$ is the metric on the $T^4$.   

The harmonic functions in \eqref{ns5f1} are given by
\begin{eqnarray}\label{harfun}
f=1-\frac{r_0^2}{r^2}~, \ \ \ f_{1,5,n}=1+\frac{r_{1,5,n}^2}{r^2}~, \ \ \ r_{1,5,n}^2=r_0^2\sinh^2\alpha_{1,5,n}~.
\end{eqnarray}
The parameters $\alpha_{1,5,n}$  are related to $k,p,n$ by
\begin{eqnarray}\label{charges}
\sinh2\alpha_1=\frac{2l_s^2g^2p}{r_0^2 v}~, \ \ \ \sinh2\alpha_5=\frac{2l_s^2k}{r_0^2}~, \ \ \ \sinh2\alpha_n=\frac{2l_s^4g^2n}{r_0^2R^2v}~,
\end{eqnarray}
where $v$ is related to the volume of $T^4$ as 
\begin{eqnarray}\label{vt4}
V_{T^4}=(2\pi)^4l_s^4v~.
\end{eqnarray}

Far away from the NS5 branes (\ie\ $r\to\infty$) the geometry becomes asymptotically flat (\ie\ $\mathbb{R}^{1,9}$). See figure \ref{spacetime} for a pictorial illustration.

To obtain the decoupled theory on the NS5 branes, one takes the limit $g\to 0$ and concentrates on radial distances $r$ of the order $gl_s$ such that the quantity $g^2l_s^2/r_0^2$ is held fixed. In this limit $f_5$ takes the form
\begin{eqnarray}\label{f5}
f_5=	\frac{kl_s^2}{r^2} \ \ \implies \ r_5^2=kl_s^2~.
\end{eqnarray}
This is equivalent to taking the near horizon limit of the NS5 branes by dropping the $1$ from the harmonic function $f_5$. In this decoupling limit, the quantities $\alpha_{1,n}$ remain finite.

In the decoupling limit, the background geometry takes the following form
\begin{eqnarray}\label{nhns5f1}
\begin{split}
&ds^2=\frac{1}{f_1}\left[-\frac{f}{f_n}dt^2+f_n\left(dx-\frac{r_0^2\sinh2\alpha_n}{2f_nr^2}dt\right)^2\right]+\frac{kl_s^2}{fr^2}dr^2+\underbrace{kl_s^2d\Omega_3^2+ds^2_{T^4}}_{S^3\times T^4}~,\\
&H=dx\wedge dt\wedge d\left(\frac{r_0^2\sinh2\alpha_1}{2f_1r^2}\right)+r_0^2\sinh2\alpha_5d\Omega_3~,\\
& e^{2\Phi}=g^2\frac{kl_s^2}{f_1r^2}~.
\end{split}
\end{eqnarray}
The background \eqref{nhns5f1} describes the region shaded in red in figure \ref{spacetime}. One can further take the limit of going to the near horizon geometry of the F1 strings by simply dropping the $1$ from the harmonic function $f_1$ of the F1 strings. In this limit, the geometry is that of arotating BTZ black hole. In figure \ref{spacetime}, this regime of the geometry (namely the $AdS_3$/BTZ regime) is denoted by the round bottom part of the figure.  For the rest of the discussion let us compactify on $S^3\times T^4$ and drop the contribution coming from this part of the ten-dimensional background.

To relate \eqref{nhns5f1} to the background \eqref{bkg2} let us consider the following change of coordinates:
\begin{eqnarray}\label{rho}
\rho^2=r^2+r_0^2\sinh^2\alpha_n~.
\end{eqnarray}
In this new coordinate system, the three-dimensional background \eqref{nhns5f1} (excluding $S^3\times T^4$) takes the following form
\begin{eqnarray}
\begin{split}\label{blackstring}
&ds^2=-\frac{(\rho^2-\rho_-^2)(\rho^2-\rho_-^2)}{\ell^2\rho^2}dt^2+\frac{kl_s^2\rho^2}{(\rho^2-\rho_-^2)(\rho^2-\rho_-^2)}d\rho^2+\frac{\rho^2}{\ell^2}\left(dx-\frac{\rho_+\rho_-}{\rho^2}dt\right)^2~,\\
&H= dx\wedge dt\wedge d\left(\frac{r_0^2\sinh2\alpha_1}{2\ell^2}\right)~,\\
& e^{2\Phi}=g^2\frac{kl_s^2}{\ell^2}~,
\end{split}
\end{eqnarray}
where 
\begin{eqnarray}\label{rhopm}
\rho_+^2=r_0^2\cosh^2\alpha_n~, \ \ \ \ \rho_-^2=r_0^2\sinh^2\alpha_n~,
\end{eqnarray}
and
\begin{eqnarray}\label{newell}
\ell^2=\rho^2-r_0^2\sinh^2\alpha_n+r_0^2\sinh^2\alpha_1~.
\end{eqnarray}

The background \eqref{blackstring} with \eqref{rhopm} and \eqref{newell} can be precisely identified with \eqref{bkg2} with \eqref{horizons} and \eqref{ell} once we make the following identifications:
\begin{eqnarray}
\begin{split}
&r_0^2=2kl_s^2\cos\chi\cos(\chi-\psi)~,\\
&\sinh^2\alpha_1-\sinh^2\alpha_n=\tan\chi\tan(\chi-\psi)~,\\
&\sinh2\alpha_1=\frac{2\sin(\chi-\psi/2)\cos(\psi/2)}{\cos\chi\cos(\chi-\psi)}~,\\
&g=e^{\Phi_0}~.
\end{split}
\end{eqnarray}

\section{BRST currents, physical vertex operators and the spectrum}\label{sec4}

In this section, we perform  BRST quantization of the coset description discussed in section \ref{sec3}, define BRST invariant physical worldsheet vertex operators and eventually calculate the spectrum of the spacetime theory. We will show that the spectrum for the winding one sector agrees exactly with the spectrum of a $T\bar{T}$ deformed CFT$_2$. 

\subsection{BRST quantization}

The gauging \eqref{gauget} of the $U(1)$ subgroup of $SL(2,\mathbb{R})\times U(1)$ can be effectuated by introducing a non-dynamical $U(1)$ gauge field $A,\bar{A}$ and adding corresponding gauge invariant terms to the action. In this subsection, we aim to construct a gauge fixed path integral using Faddeev-Popov ghost fields and calculate the BRST current of the $\frac{SL(2,\mathbb{R})_k\times U(1)}{U(1)}$ coset CFT.  We closely follow that the techniques of BRST quantization of a gauged WZW models adapted in \cite{Karabali:1989dk}. 

The gauge fields in the gauged action \eqref{gwzw2} are non-dynamical: it doesn't contain kinetic terms associated with the gauge fields. Therefore they act as Lagrange multipliers that set the gauge currents to zero. 
Quantum mechanically one needs to path integrate over all possible gauge inequivalient field configuration of the $A,\bar{A}$. The path integral of the theory is given by
\begin{eqnarray}\label{Z1}
Z=\int [Dg][Dx][DA][D\bar{A}]e^{-S_g[g,x,A,\bar{A}]}~,
\end{eqnarray}
where $S_g[g,x,A,\bar{A}]$ is given in \eqref{gwzw2}.

It has been shown in \cite{Karabali:1989dk} that gauging the original WZW model will turns on an additional $U(1)$ \footnote{Loosely speaking one can think of this $U(1)$ as the $U(1)$ in the denominator of the coset $\frac{SL(2,\mathbb{R})_k\times U(1)}{U(1)}$.} WZW term, $S[w]$, given by  
\begin{eqnarray}\label{Sw}
S[w]=\frac{1}{2\pi}\int_\Sigma d^2z \partial w\bar{\partial} w~.
\end{eqnarray}
along with the ghost sector $S_{gh}$ \eqref{Sghost}. The new WZW term, \eqref{Sw},  comes with the ``wrong sign'' in the path integral \eqref{Z3}. This is also reflected in the $w(z_1)w(z_2)$ OPE
\begin{eqnarray}\label{wwope}
w(z_1)w(z_2)=\frac{1}{2}\ln (z_1-z_2)~.
\end{eqnarray}
The corresponding $U(1)$ currents 
\begin{eqnarray}
J_w=i\partial w~, \ \ \ \ \ \bar{J}_w=i\bar{\partial} w~,
\end{eqnarray}
satisfy the following OPE algebra
\begin{eqnarray}
J_w(z_1)J_w(z_2)=-\frac{1/2}{(z_1-z_2)^2}~.
\end{eqnarray}
The field $w$ can be thought of as the coordinate that parametrizes the $U(1)$ subgroup of $SL(2,\mathbb{R})\times U(1)$  that we want to gauge and the gauge fields $A,\bar{A}$ are related to $w$ in the following way:
  \begin{eqnarray}\label{hhbar}
A=\partial h h^{-1}=-\partial v~, \ \ \ \ \bar{A}=\bar{\partial}\bar{h}\bar{h}^{-1}=-\bar{\partial} u ~,
\end{eqnarray}
where $h,\bar{h}$ are independent group elements of the gauge group $U(1)$ given by
\begin{eqnarray}\label{hhbarw}
h=e^{-v}, \ \ \ \ \bar{h}=e^{-u}, \ \  \text{ and } \ \  w=u-v~.
\end{eqnarray}
The path integrals over $A,\bar{A}$ can be replaced by integrals over the fields $u,v$ with appropriate Jacobian 
factor in the measure: \footnote{Note that in the case of non-Abelian gauging, the determinants of the usual derivatives are replaced by the determinants of the covariant derivatives.}
\begin{eqnarray}\label{measure}
[DA]\to [Du]\det(\partial)~, \ \ \ \ [D\bar{A}]\to [Dv]\det(\bar{\partial})~.
\end{eqnarray}
The functional integral representation of the determinant of the holomorphic and anti-holomorphic derivatives are obtained by introducing two sets of the Faddeev-Popov $b,c$ and $\bar{b},\bar{c}$ ghosts:
\begin{eqnarray}\label{FPghosts}
\det(\partial)\det(\bar{\partial})=\int[Db][Dc][D\bar{b}][D\bar{c}]e^{-\frac{1}{2\pi}\int_\Sigma d^2z(b\bar{\partial}c+\bar{b}\partial\bar{c})}~.
\end{eqnarray}
The ghost fields are normalized such that 
\begin{eqnarray}\label{bcope}
c(z_1)b(z_2)\sim\frac{1}{z_1-z_2}+\cdots~, \ \ \ \  \bar{c}(\bar{z}_1)\bar{b}(\bar{z}_2)\sim\frac{1}{\bar{z}_1-\bar{z}_2}+\cdots~.
\end{eqnarray}
Substituting \eqref{measure},\eqref{FPghosts} in \eqref{Z1} and following same line of arguments as in \cite{Karabali:1989dk} one obtains
\begin{eqnarray}\label{Z3}
Z=\int[Dg][Dx][Du][Dv][Db][Dc][D\bar{b}][D\bar{c}]e^{-S[g]-S[x]+S[w]-S_{gh}}~,
\end{eqnarray}
where the ghost action $S_{gh}$ is given by
\begin{eqnarray}\label{Sghost}
S_{gh}=\frac{1}{2\pi}\int_\Sigma d^2z(b\bar{\partial}c+\bar{b}\partial\bar{c})~.
\end{eqnarray}

Next let us fix gauge: $\bar{A}=0$. This is equivalent to setting $u$ equal to some constant. Thus the path integral \eqref{Z3} takes the following form
\begin{eqnarray}
Z=\int[Dg][Dx][Dw][Db][Dc][D\bar{b}][D\bar{c}]e^{-S[g]-S[x]+S[w]-S_{gh}}~.
\end{eqnarray}
The total gauge fixed action is given by
\begin{eqnarray}\label{stotal}
S_{tot}=S[g]+S[x]-S[w]+S_{gh}~.
\end{eqnarray}

The variation of $S_{tot}$ \eqref{stotal} under the infinitesimal changes (holomorphic) $\delta  g, \delta x, \delta w, \delta b, \delta c $ is
\begin{eqnarray}\label{dstot}
\delta S_{tot}=-\frac{1}{2\pi}\int_\Sigma d^2z~ \left[k\tr\left(\partial gg^{-1}\bar{\partial}(g^{-1}\delta g)\right)-\partial x\bar{\partial} \delta x+2\partial w\bar{\partial} \delta w-\delta b\bar{\partial} c-b\bar{\partial}\delta c\right].
\end{eqnarray}
The BRST transformations of the fields $g,x,w,b,c$ are given by \cite{Bastianelli:1990ey}
\begin{eqnarray}\label{deltab}
\delta_B g=\eta c g \sigma_3~, \ \ \ \delta_B x= \eta c x~, \ \ \ \delta_B w=\eta c~, \ \ \ \delta_B b=\eta({\bf{J}}+2 J_w)~, \ \ \ \delta_Bc=0~,
\end{eqnarray}
where $\eta$ is an arbitrary Grassmann parameter. Thus the change in the total gauge fixed action under the infinitesimal transformations \eqref{deltab}, takes the form
\begin{eqnarray}\label{deltabstot}
\delta_B S_{tot}=\frac{1}{2\pi}\int_\Sigma d^2z~(\bar{\partial}\eta) c({\bf{J}}+2J_w)~.
\end{eqnarray}
From \eqref{deltabstot} one can read off the BRST current $J_{BRST}$ as
\begin{eqnarray}\label{jbrst}
J_{BRST}=c({\bf{J}}+2J_w)~.
\end{eqnarray}
Switching the chiralities one can similarly define the anti-holomorphic component of the BRST current $\bar{J}_{BRST}$.

The corresponding BRST charges are given by
\begin{eqnarray}\label{qbrst}
Q_{BRST}=\frac{1}{2\pi i}\oint dz~ J_{BRST}~, \ \ \ \ \ \bar{Q}_{BRST}=\frac{1}{2\pi i}\oint d\bar{z}~ \bar{J}_{BRST}~.
\end{eqnarray}
The BRST charges \eqref{qbrst} are nilpotent: $Q_{BRST}^2=\bar{Q}_{BRST}^2=0$. This follows from the fact that 
\begin{eqnarray}\label{brst2ptfn}
\langle ({\bf{J}}+2J_w)(z_1)({\bf{J}}+2J_w)(z_2)\rangle=0~.
\end{eqnarray}
Thus the physical states of the Hilbert space of the coset CFT are states that are annihilated by $Q_{BRST},\bar{Q}_{BRST}$  (\ie\ BRST closed)
\begin{eqnarray}\label{phys}
Q_{BRST} |{\rm{phys}}\rangle=\bar{Q}_{BRST}|{\rm{phys}}\rangle=0~,
\end{eqnarray}
and are defined up to BRST exact states.

In the discussion that follows, we will impose the BRST constraints on the vertex operators and eventually compute the spectrum.

\subsection{Physical vertex operators and spectrum}

\subsubsection{BRST invariant vertex operators}
In this subsection, we will construct the physical worldsheet vertex operators of ten-dimensional type II  critical string theory on $\frac{SL(2,\mathbb{R})_k\times U(1)}{U(1)}\times \mathcal{N}_7$ where $\mathcal{N}_7$ is a seven-dimensional spacelike compact manifold. Criticality of the worldsheet theory demands that the total matter central charge of the worldsheet CFT must add up to $15$. This implies
\begin{eqnarray}\label{ccn}
c_{coset}+\frac{3}{2}+c_\mathcal{N}=15~,
\end{eqnarray}
where $c_{\mathcal{N}}$ of the worldsheet central charge on $\mathcal{N}_7$ and $c_{coset}$ is the central charge of the coset CFT  $\frac{SL(2,\mathbb{R})_k\times U(1)}{U(1)}\times $ given by
\begin{eqnarray}\label{ccads}
c_{coset}=c_{SL(2,\mathbb{R})\times U(1)}-c_{U(1)}=3+\frac{6}{k}~,
\end{eqnarray}
which, as expected, is also the central charge of a WZW model on $SL(2,\mathbb{R})$ at level $k+2$ \footnote{In superstring theory the currents $J^A$ get contribution from the worldsheet  bosons and fermions. Thus the level of the $SL(2,\mathbb{R})$ current algebra in superstring theory is $k+2$ where $k$ is the level of just the bosonic sector.}.

A primary vertex operator of the worldsheet theory can be written as a product of a primary operator on $\frac{SL(2,\mathbb{R})_k\times U(1)}{U(1)}$ times a primary operator on $\mathcal{N}_7$. Suppose $\Psi(z,\bar{z})$ be a vertex operator on $SL(2,\mathbb{R})_k\times U(1)_x$, then a vertex operator on $\frac{SL(2,\mathbb{R})_k\times U(1)_x}{U(1)_w}$ is given by
\begin{eqnarray}\label{vt}
V_t(z,\bar{z})=\Psi(z,\bar{z}) e^{il_s(\kappa_Lw_L+\kappa_Rw_R)}~,
\end{eqnarray}
where $w(z,\bar{z})=w_L(z)+w_R(\bar{z})$ \footnote{Note that the vertex operator of the coset $G/H$ can be expressed as the product of a vertex operator coming from $G$ and a vertex operator coming from $H$. This follows from  the fact that the gauge fixed action can be written as $S_G-S_H+S_{ghost}$. Of course the vertex operator constructed in this way is not necessarily gauge invariant. Thus one needs to impose BRST invariance on the vertex operators to construct states in the BRST cohomology.}. Note that above we have introduced the suffix $x$ and $w$ on the group $U(1)$ to remind the reader that $x$ parametrizes the $U(1)_x$ in the numerator of the coset and $w$ parametrized $U(1)_w$ in the denominator of the coset. Since $U(1)_w$ is non-compact, there are no winding. Thus $\kappa_L=\kappa_R=\kappa$ is a continuous variable. Low-lying vertex operator $\Psi(z,\bar{z})$  on $SL(2,\mathbb{R})_k\times U(1)_x$ can be expressed as 
\begin{eqnarray}\label{phi}
\Psi(z,\bar{z})=V^j_{m,\bar{m}} e^{il_s(p_Lx_L+p_Rx_R)}~,
\end{eqnarray}
where $V^j_{m,\bar{m}}$'s are primary vertex operators in the continuous representation (\ie\ $j+\frac{1}{2}\in i\mathbb{R}$) of Euclidean $SL(2,\mathbb{R})$($ \equiv H_3^+$) obtained by diagonalizing $J^3,\bar{J}^3$ with eigenvalues $im,-i\bar{m}$ (see \eg\ section 2.2 of \cite{Elitzur:2002rt}). The charges $p_{L,R}$ are the left and right moving $U(1)_x$ momentum charges. Since $U(1)_x$ is compact,  
$p_{L,R}$  can be written as
\begin{eqnarray}\label{plpr}
p_{L,R}=\frac{n}{R}\pm \frac{\omega R}{l_s^2}~,
\end{eqnarray}
where $n\in\mathbb{Z}$ is the discrete momentum number and $\omega$ is the winding quantum number.

Let $V(z,\bar{z})$ be an operator on the $\frac{SL(2,\mathbb{R})_k\times U(1)}{U(1)}\times \mathcal{N}_7$, then $V(z,\bar{z})$ is given by
\begin{eqnarray}\label{physv}
V(z,\bar{z})=V_t(z,\bar{z})V_{\mathcal{N}}~,
\end{eqnarray}
subject to the constraint that the state dual to $V_t(z,\bar{z})$ is BRST closed (\ie\ OPE between $J_{BRST} $ and $V_t$ is regular). Here $V_{\mathcal{N}}$ is a primary vertex operator on $\mathcal{N}_7$. 
The BRST constraints are given by
\begin{eqnarray}
\begin{split}\label{brstcond}
&\frac{2m}{\sqrt{k}}\cos(\chi-\psi)-l_sp_L\sin(\chi-\psi)+l_s\kappa=0~,\\
&-\frac{2\bar{m}}{\sqrt{k}}\cos\chi+l_sp_R\sin\chi+l_s\kappa=0~.
\end{split}
\end{eqnarray}
The above two equations \eqref{brstcond} can be combined into one single constraint equation
\begin{eqnarray}\label{brstcond1}
\frac{2}{\sqrt{k}}\left(m\cos(\chi-\psi)+\bar{m}\cos\chi\right)-l_s\left(p_L\sin(\chi-\psi) - p_R\sin\chi\right)=0~.
\end{eqnarray}

\subsubsection{Spectrum}
The stress tensor of the $\frac{SL(2,\mathbb{R})_k\times U(1)}{U(1)}\times \mathcal{N}_7$ coset CFT can be obtained via Sugawara construction
\begin{eqnarray}
\begin{split}\label{ttbar}
&T=\frac{1}{k}\eta_{AB}J^AJ^B-\partial x \partial x+\partial w\partial w+T_\mathcal{N}~,\\
&\bar{T}=\frac{1}{k}\eta_{AB}\bar{J}^A\bar{J}^B-\bar{\partial} x \bar{\partial} x+\bar{\partial} w\bar{\partial} w+\bar{T}_\mathcal{N}~,
\end{split}
\end{eqnarray}
where $T_\mathcal{N},\bar{T}_\mathcal{N}$ are the stress tensors of the worldsheet CFT on $\mathcal{N}_7$.
The dimensions of the vertex operator $V(z,\bar{z})$ \eqref{physv} can be read off from its OPE with the components of the stress tensor:
\begin{eqnarray}
\begin{split}\label{TVope}
&T(z_1)V(z_2)=\frac{1}{(z_1-z_2)^2}\left(-\frac{j(j+1)}{k}+\frac{l_s^2p_L^2}{4}-\frac{l_s^2\kappa^2}{4}+\Delta_L^ \mathcal{N}\right)V(z_2)+\cdots~,\\
&\bar{T}(\bar{z}_1)V(\bar{z}_2)=\frac{1}{(\bar{z}_1-\bar{z}_2)^2}\left(-\frac{j(j+1)}{k}+\frac{l_s^2p_R^2}{4}-\frac{l_s^2\kappa^2}{4}+\Delta_R^ \mathcal{N}\right)V(\bar{z}_2)+\cdots~,
\end{split}
\end{eqnarray}
where $\Delta_{L,R}^ \mathcal{N}$ are the worldsheet dimensions of $V_\mathcal{N}$ 
and
\begin{eqnarray}\label{js}
j=-\frac{1}{2}+is~, \ \ \ \ s\in\mathbb{R}~.
\end{eqnarray}
Thus the left and right moving dimensions $(\Delta_L,\Delta_R)$ of $V(z,\bar{z})$ are given by
\begin{eqnarray}\label{deltalr}
\Delta_{L,R}=-\frac{j(j+1)}{k}+\frac{l_s^2p_{L,R}^2}{4}-\frac{l_s^2\kappa^2}{4}+\Delta_{L,R}^ \mathcal{N}~.
\end{eqnarray}
For physical vertex operators $\mathcal{V}$ in the $(-1,-1)$ picture,
\begin{eqnarray}\label{physvo}
\mathcal{V}=e^{-\varphi}e^{-\bar{\varphi}}V(z,\bar{z})~,
\end{eqnarray}
where $\varphi,\bar{\varphi}$ are the worldsheet superconformal ghosts, the mass-shell condition is given by
\begin{eqnarray}\label{onshell}
-\frac{j(j+1)}{k}+\frac{l_s^2p_{L,R}^2}{4}-\frac{l_s^2\kappa^2}{4}+\Delta_{L,R}^ \mathcal{N}=\frac{1}{2}~.
\end{eqnarray}
Adding the two equations \eqref{onshell} one can write
\begin{eqnarray}\label{ksquare}
\kappa^2=\left(\frac{n}{R}\right)^2+\left(\frac{\omega R}{l_s^2}\right)^2+\frac{2}{l_s^2}\left(-\frac{2j(j+1)}{k}+\Delta_L^ \mathcal{N}+\Delta_R^ \mathcal{N}-1\right)~.
\end{eqnarray}
The difference of the two equations \eqref{onshell} gives
\begin{eqnarray}\label{diffdelta}
\Delta_R^ \mathcal{N}-\Delta_L^ \mathcal{N}=n\omega~.
\end{eqnarray}

Next we will identify $\kappa$ as the energy of the string excitation described by the physical vertex operator \eqref{physvo}.   The asymptotic expansion (large $\theta$) of the vertex operator $V^j_{m,\bar{m}}$ in Lorentzian $AdS_3$, upon imposing the gauge fixing condition \eqref{gfixing}, is given by \cite{Elitzur:2002rt}
\begin{eqnarray}\label{vjmmbae}
V^j_{m,\bar{m}}(\theta\to\infty)=e^{-\theta+iy(m-\bar{m})}\left[e^{2is\theta}+R(j;m,\bar{m})e^{-2is\theta}\right]~,
\end{eqnarray}
where
\begin{eqnarray}
R(j;m,\bar{m})=\frac{\Gamma(j+1+im)\Gamma(j+1+i\bar{m})\Gamma(-2j-1)}{\Gamma(-j+im)\Gamma(-j+i\bar{m})\Gamma(2j+1)}~.
\end{eqnarray}
For large $\theta$, the vertex operator $V_t$ in \eqref{vt} with \eqref{phi}, upon imposing the BRST constraint \eqref{brstcond} takes the following expected form
\begin{eqnarray}\label{vtae}
V_t(\theta\to\infty)=e^{-\theta}\left[e^{2is\theta}+R(j;m,\bar{m})e^{-2is\theta}\right]e^{-i\kappa T}e^{i(p_LX_L+p_RX_R)}~,
\end{eqnarray}
where $T$ is a timelike direction and $X=X_L+X_R$ is a spacelike direction of the spacetime CFT, given by
\begin{eqnarray}\label{TXLR}
\begin{split}
T&=t-l_sw~,  \\
 X_L&=l_sx_L+\frac{\cos\chi\sin(\chi-\psi)}{2\cos(\chi-\psi/2)\cos(\psi/2)}t~, \\ 
 X_R&=l_sx_R-\frac{\sin\chi\cos(\chi-\psi)}{2\cos(\chi-\psi/2)\cos(\psi/2)}t~.
\end{split}
\end{eqnarray} 
That $X=X_L+X_R$ is a spacial direction of the spacetime theory can also be verified by taking the asymptotic limit (\ie\ $\theta\to\infty$) of the background metric  \eqref{bkg1} with the appropriate coordinate transformations \eqref{coortrfn}.

As stated above, from \eqref{vtae} one can identify $\kappa$ as the energy of the string excitation considered in  \eqref{physvo}. The energy $E$ of the string with winding $\omega$ above the BPS configuration is given by
\begin{eqnarray}\label{eng}
E=\kappa-\frac{\omega R}{l_s^2}~.
\end{eqnarray}
Similarly from \eqref{vtae} one can also identify $p_{L,R}$ as the left and right moving momentum of the spacetime theory.

In massless BTZ, the dispersion relation in the winding sector $\omega$ is given by \cite{Parsons:2009si}
\begin{eqnarray}\label{spectads}
E_{L,R}=\frac{1}{\omega}\left[-\frac{j(j+1)}{k}+\Delta_{L,R}^ \mathcal{N}-\frac{1}{2}\right]~.
\end{eqnarray}
States whose spectrum is given by \eqref{spectads} can be thought of as strings with winding $\omega$ around the spacial circle in BTZ geometry with a certain radial momentum given by
\begin{eqnarray}\label{radialmom}
p_{\phi}=s\sqrt{\frac{2}{k}}~,
\end{eqnarray}
where $s$ is defined in \eqref{js} and in a particular state of transverse oscillation. 

It has been pointed out in \cite{Giveon:2005mi} that the spectrum \eqref{spectads} is also the spectrum of the $\mathbb{Z}_\omega$ twisted sector of a symmetric product CFT $\mathcal{M}^N/S_N$, where the block CFT $\mathcal{M}$ has central charge $c_\mathcal{M}=6k$ associated with a single long string in the BTZ geometry with winding $\omega$. Such long strings with spectrum \eqref{spectads}, correspond to an operator of dimension $h_{\omega}$ in the symmetric orbifold $\mathcal{M}^N/S_N$ where the spectrum of states in the Ramond sector above the Ramond vacuum is given by \cite{Argurio:2000tb}
\begin{eqnarray}\label{spectsymo}
E_L=h_\omega-\frac{k\omega}{4}~, \ \ \ \ E_{R}=\bar{h}_\omega-\frac{k\omega}{4}~.
\end{eqnarray}
Equating \eqref{spectads} and \eqref{spectsymo} one can write 
\begin{eqnarray}\label{homega}
h_\omega=\frac{h_1}{\omega}+\frac{k}{4}\left(\omega-\frac{1}{\omega}\right)~ ,
\end{eqnarray}
which relates the dimensions of the $\mathbb{Z}_\omega$ twisted sector, $h_\omega$, to the dimensions of operators of the untwisted sector $h_1$.

Substituting \eqref{spectads},\eqref{spectsymo} and \eqref{homega} in  \eqref{ksquare} and \eqref{eng}, one obtains
\begin{eqnarray}\label{spect1}
\left(E+\frac{\omega R}{l_s^2}\right)^2-\left(\frac{\omega R}{l_s^2}\right)^2=\frac{2}{l_s^2}\left(h_1+\bar{h}_1-\frac{k}{2}\right)+\left(\frac{n }{R}\right)^2,
\end{eqnarray}
and 
\begin{eqnarray}\label{hhbardiff}
\bar{h}_1-h_1=n~.
\end{eqnarray}
Considering the winding $\omega =1$ sector which also corresponds to the untwisted sector of the spacetime symmetric orbifold theory, \eqref{spect1} can be cast into the following form
\begin{eqnarray}\label{ttbarspect}
E=-\frac{R}{l_s^2}+\sqrt{\left(\frac{R}{l_s^2}\right)^2+\frac{2}{l_s^2}\left(h_1+\bar{h}_1-\frac{c_{\mathcal{M}}}{2}\right)+\left(\frac{h_1-\bar{h}_1}{R}\right)^2}~,
\end{eqnarray} 
 where we considered only the positive branch of the square root. This can be compared to the deformed spectrum obtained upon irrelevant deformation of the form $\delta\mathcal{L}=-tT\bar{T}$ of a CFT$_2$ \cite{Smirnov:2016lqw,Cavaglia:2016oda} or the single trace $T\bar{T}$ deformation of string theory in $AdS_3$ in the untwisted sector \cite{Giveon:2017nie,Giveon:2017myj,Chakraborty:2019mdf} once we identify the irrelevant coupling $t$ as $t=l_s^2$. For $\omega>1$, \eqref{spect1} is the spectrum of the $\mathbb{Z}_w$ twisted sector of single trace $T\bar{T}$ deformation of string theory in $AdS_3$.

\section{Correlation functions}\label{sec5}

In the previous section, we have established the fact that the coset CFT $\frac{SL(2,\mathbb{R})_k\times U(1)}{U(1)}$ with the gauge currents given by \eqref{gaugecurr} describes the full near horizon theory of the NS5 branes  of a system of $k$ NS5 branes and $p$ F1 strings with momentum modes along the direction on which wraps the F1 strings at all temperatures. We also argued that at zero temperature, one would obtain the same background upon single trace $T\bar{T}$ deformation of string theory in $AdS_3$ with only the NS-NS H-flux turned on. We showed that the spectrum obtained from the worldsheet approach agrees precisely with that obtained in the case of single trace $T\bar{T}$ deformed string theory in $AdS_3$. In this section, as a check of what had been argued in the previous sections, we would like to compute the two-point functions of the two local operators of the spacetime theory from the worldsheet approach (using the fact that the worldsheet sigma model is given by $\frac{SL(2,\mathbb{R})_k\times U(1)}{U(1)}\times \mathcal{N}_7$) and show that it matches precisely with the one computed from the supergravity approach.

\subsection{Worldsheet approach}\label{sec5.1}

A large class of observables in string theory in $AdS_3$ are given by vertex operators, which in the $(-1,-1)$ picture number sector take the form
\begin{eqnarray}\label{O(x)}
\mathcal{O}(x)=\int d^2z e^{-\varphi-\bar{\varphi}}\Phi_{h}(z;x)V_{\mathcal{N}}(z)~,
\end{eqnarray}
where as before $V_{\mathcal{N}}$ is an $\mathcal{N}=1$ superconformal primary of the worldsheet CFT on $\mathcal{N}_7$ with worldsheet dimension $(\Delta_L^\mathcal{N},\bar{\Delta}_R^\mathcal{N})$, $\Phi_{h}(z;x)$ are the usual vertex operators on $AdS_3$ and labelled by the boundary position $x$ and the worldsheet insertion point $z$ and has spacetime dimension $(h,\bar{h})$\footnote{The operator $\Phi_{h,\bar{h}}(z;x)$ in this section is related to the operator $V^j_{m,\bar{m}}(z,\bar{z})$ in the previous section by Mellin transformation. For example, in Euclidean $AdS_3$ they are related as follows:
\begin{eqnarray}\label{vphi}
V^j_{m,\bar{m}}(z,\bar{z})=\int d^2x~ x^{j+im}\bar{x}^{j-i\bar{m}}\Phi_{h,\bar{h}}(z,\bar{z};x,\bar{x})~.
\end{eqnarray}}. 
For simplicity we assume $h=\bar{h}$ and $\Delta_L^\mathcal{N}=\bar{\Delta}_R^\mathcal{N}=\Delta_\mathcal{N}$. The spacetime dimension $h$ of $\Phi_{h}(z;x)$  is related to the $SL(2,\mathbb{R})$ quantum number $j$ via
\begin{eqnarray}
h=j+1~.
\end{eqnarray}
The operator $\mathcal{O}(x)$ \eqref{O(x)} is a local operator in the spacetime theory that satisfies the on-shell condition
\begin{eqnarray}\label{onshelld}
\Delta_h+\Delta_\mathcal{N}-\frac{1}{2}=0~,
\end{eqnarray}
where 
\begin{eqnarray}\label{deltah}
\Delta_h=-\frac{h(h-1)}{k}
\end{eqnarray}
is the worldsheet dimension of $\Phi_h(z;x)$.

The operator $V_{\mathcal{N}}$, coming from the internal worldsheet CFT $\mathcal{N}_7$, is normalized such that 
\begin{eqnarray}\label{vnvn}
\langle V_{\mathcal{N}}(z_1)V_{\mathcal{N}}(z_2)\rangle = \frac{1}{|z_1-z_2|^{4\Delta_{\mathcal{N}}}}~.
\end{eqnarray}
The worldsheet operators $\Phi_h(z;x)$ are normalized such that 
\begin{eqnarray}\label{phi2pt}
\langle \Phi_{h}(z_1;x_1)  \Phi_{h'}(z_2;x_2)  \rangle=\delta(h-h')\frac{\frac{2}{\pi}(2h-1) X^{2h-1}}{|z_1-z_2|^{4\Delta_h}|x_1-x_2|^{4h}}~,
\end{eqnarray}
where $X$ is an arbitrary constant whose value can be adjusted by shifting the radial coordinate or rescaling the transverse field theory coordinates.
Note that the choice of this particular normalization of the operators in \eqref{phi2pt} is non-standard and soon it will become clear the reason behind the choice of this particular normalization. The two-point  function \eqref{phi2pt} in momentum space can be expresses as \cite{Kutasov:1999xu}
\begin{eqnarray}\label{phi2ptmom}
\langle \Phi_{h}(z_1;p)  \Phi_{h'}(z_2;-p)  \rangle= \delta(h-h')2 X^{2h-1}\frac{\Gamma(1-2h)}{\Gamma(2h-1)}\left(\frac{p^2}{4}\right)^{2h-1}\frac{1}{|z_1-z_2|^{4\Delta_h}}~.
\end{eqnarray} 

Using the worldsheet two-point function \eqref{phi2ptmom} one can construct the two-point function of the boundary theory in momentum space \cite{Giveon:2001up,Maldacena:2001km}
\begin{eqnarray}\label{oo2ptmom}
\langle \mathcal{O}(p)\mathcal{O}(-p)\rangle=2(2h-1)X^{2h-1}\frac{\Gamma(1-2h)}{\Gamma(2h-1)}\left(\frac{p^2}{4}\right)^{2h-1}.
\end{eqnarray}
Fourier transforming \eqref{oo2ptmom} one obtains \footnote{In writing the position space two-point function we were not very careful about the normalization.}
\begin{eqnarray}\label{ox1ox2}
\langle \mathcal{O}(x_1)\mathcal{O}(x_2)\rangle \sim \frac{1}{|x_1-x_2|^{4h}}~.
\end{eqnarray}

The operator $\mathcal{O}(x)$ is a local operator in the boundary CFT  dual to string theory in $AdS_3$. In the coset construction, the spacetime theory is no longer conformal. We are interested in a local operator in the spacetime theory that can be expressed (in momentum space) as
\begin{eqnarray}
\mathcal{O}(p)=\int d^2z e^{-\varphi-\bar{\varphi}}\Phi_{h_p}(z;p) e^{il_s(p_Lx_L+p_Rx_R)} e^{il_s\kappa w}V_{\mathcal{N}}(z)~,
\end{eqnarray}
subject to the BRST constraint  \eqref{brstcond}. The dimension of the operator $\mathcal{O}(p)$ in the deformed theory \footnote{By deformed theory we mean the coset theory $\frac{SL(2,\mathbb{R})_k\times U(1)}{U(1)}$.}, which we denote by $h_p$  satisfies the condition \eqref{onshell} \footnote{Remember that $t=l_s^2$.}
\begin{eqnarray}\label{hhp}
&& (2h_p-1)=\sqrt{(2h-1)^2+tkp^2}~,
\end{eqnarray}
where for simplicity let us assume $p_L=p_R$ and $\Delta_L^\mathcal{N}=\Delta_R^\mathcal{N}=\Delta_\mathcal{N}$ and we denote
\begin{eqnarray}
p^2=p_{L,R}^2-\kappa^2~,
\end{eqnarray}
as the square of the norm of the two-dimensional energy-momentum vector of the spacetime theory.
The two-point function of the operator $\mathcal{O}$ in the coset theory turns out to be given by the same expression \eqref{oo2ptmom} with $h$ replaced by $h_p$
\begin{eqnarray}\label{doo2ptmom}
\langle \mathcal{O}(p)\mathcal{O}(-p)\rangle=2(2h_p-1)X^{2h_p-1}\frac{\Gamma(1-2h_p)}{\Gamma(2h_p-1)}\left(\frac{p^2}{4}\right)^{2h_p-1}.
\end{eqnarray}
The two-point function \eqref{doo2ptmom} agrees with the one calculated in \cite{Asrat:2017tzd,Giribet:2017imm} by a related  approach. The two-point function \eqref{doo2ptmom} has an intricate pole structure that has been discussed in details in \cite{Asrat:2017tzd}.  

Another important point to note is that the part $(p^2)^{2h_p-1}$  in the two-point function \eqref{doo2ptmom} is possibly the universal sector because the rest of the terms in \eqref{doo2ptmom} excluding $(p^2)^{2h_p-1}$ can be absorbed in the normalization of the operators $\Phi_h(z;x)$ in string theory in $AdS_3$. The origin of this ambiguity in the momentum dependent normalization has to do with the choice of contact terms. This issue has been discussed in details in  \cite{Asrat:2017tzd}.

In the next subsection, we will reproduce the same two-point function \eqref{doo2ptmom} of the deformed theory from supergravity analysis.

\subsection{Supergravity approach}\label{sec5.2}

In the discussion that follows, we calculate the correlation function of two local operators in the  dual spacetime theory starting from supergravity. We will show that the two-point function thus calculated agrees precisely with the one computed from the worldsheet approach in the previous subsection.  

\subsubsection{The supergravity background at zero temperature}

Let us start with the type IIB supergravity action in three dimensions:
\begin{eqnarray}\label{sugraact}
S_{{\rm IIB}}= \int d^3 X \sqrt{-g}e^{-2\Phi}\left( R+4g^{\mu\nu}\partial_\mu\Phi\partial_\nu\Phi-\frac{1}{12}H^2-4\Lambda\right)~.
\end{eqnarray}
A solution to the supergravity equations of motion is the background \eqref{nhns5f1}. For the purpose of calculating the correlation functions of the spacetime theory and compare with the one calculated from the worldsheet approach, we will set $r_0=r_n=0$ in \eqref{nhns5f1}. The background thus obtained is given by
\begin{eqnarray}\label{nearns5}
\begin{split}
ds^2&=- \frac{1}{f_1}dt^2+\frac{1}{f_1}dx^2+f_5r^2dr^2~,\\
e^{2\Phi}&=e^{2\Phi_0}\frac{f_5}{f_1}~,\\
B_{tx}&=\frac{1}{f_1}~,
\end{split}
\end{eqnarray}
where we remind the reader that
\begin{eqnarray}\label{f1f5}
f_1=\lambda+\frac{r_1^2}{r^2}~, \ \ \ f_5=\frac{r_5^2}{r^2}=\frac{kl_s^2}{r^2}~.
\end{eqnarray}
Note that in the harmonic function $f_1$ we have introduced the marginal worldsheet deformation parameter $\lambda$. Without loss of generality, one can set $\lambda=1$ \cite{Giveon:2017nie}. In this subsection, we prefer to keep the parameter $\lambda$  in our calculation as it may be helpful in comparing the result from the supergravity approach with that obtained from the worldsheet approach.  For the rest of the discussion in this paper we will restrict ourselves to $\lambda>0$. The background \eqref{nearns5}  is the near horizon geometry of a stack of NS5 branes of a system of $k$ NS5 branes and $p$ F1 strings at zero temperature.\footnote{The $r_1^2$ in the harmonic function $f_1$ is given by $r_1^2=\frac{e^{2\Phi_0}l_s^2p}{v}$ where $v$ is related to the volume of the $T^4$ defined in \eqref{vt4}.}  As stated before, the background $\mathcal{M}_3$ interpolates between flat spacetime with a linear dilaton in the UV to $AdS_3$ in the IR.

Let us perform the following change of coordinate:
\begin{eqnarray}
z=\frac{r_5^2}{r}~.
\end{eqnarray}
The background metric in this new coordinate system takes the following form:
\begin{eqnarray}
ds^2&=&\frac{r_5^2}{z^2}\left( \frac{r_5^2}{ l^2} (-dt^2+dx^2) +dz^2\right)~,
\end{eqnarray}
where 
\begin{eqnarray}
l^2=	\frac{\lambda r_5^4}{z^2}+r_1^2~.
\end{eqnarray}
In this new coordinate system, $z\to0$ is the boundary of spacetime.
Note that the coordinates $t,z,x$ have dimensions of length and the parameter $\lambda$ is dimensionless.

\subsubsection{Massive scalar field minimally coupled to $\mathcal{M}_3$}

Let us consider a scalar field $\Psi$ of mass $m$ minimally coupled to the background \eqref{nearns5}. The supergravity action then takes the form \footnote{One can also add to the supergravity action a polynomial potential of the scalar field $\Psi$. In the semiclassical limit (\ie\ large $N$ and large 't Hooft), such terms are suppressed. }
\begin{eqnarray}
S &=& \int d^3 X \sqrt{-g}e^{-2\Phi}\left( R+4g^{\mu\nu}\partial_\mu\Phi\partial_\nu\Phi-\frac{1}{12}H^2-4\Lambda-\frac{1}{2}\left(g^{\mu\nu}\partial_\mu\Psi\partial_\nu\Psi+m^2\Psi^2\right)\right)~.\nonumber\\
\end{eqnarray} 
 The Klein-Gordon equation of motion of the massive scalar field is given by
\begin{eqnarray}\label{KGeqn}
\frac{e^{2\Phi}}{\sqrt{-g}}\partial_\mu\left(\sqrt{-g}e^{-2\Phi}g^{\mu\nu}\partial_\nu\Psi\right)-m^2\Psi=0~.
\end{eqnarray}
In the dual field theory living on the boundary of $\mathcal{M}_3$, this corresponds to adding a deformation of the form $\int d^2x J(x)\mathcal{O}(x)$ to the Lagrangian, where $\mathcal{O}(x)$ is an operator of dimension $\Delta$ and $J(x)$ is its source.

In the discussion that  follows, let us switch to Euclidean signature for convenience.
The bulk metric \eqref{nearns5} is invariant under translation in $t$ and $x$. This implies that the solution to the Klein-Gordon equation \eqref{KGeqn} takes the form
\begin{eqnarray}\label{sol1}
\Psi(t,z, x)=\psi(z) e^{-i\omega t}e^{iqx}~.
\end{eqnarray}
Substituting \eqref{sol1} in \eqref{KGeqn}  one obtains
\begin{eqnarray}\label{eom}
z^2\psi''(z)-z\psi'(z)-\left(m^2r_5^2+\lambda r_5^2p^2+\frac{r_1^2}{r_5^2}p^2z^2\right)\psi(z)=0~,
\end{eqnarray}
where 
\begin{eqnarray}\label{emom}
p^2=q^2+\omega^2~.
\end{eqnarray}
Imposing regularity at $z\to\infty$, the full solution to \eqref{eom} takes the form
\begin{eqnarray}\label{sol}
\psi(z)&=& Xz K_\nu(Xz)~,
\end{eqnarray}
where $K_n(x)$ is the modified Bessel function of second kind of order $n$ and argument $x$ and  
\begin{eqnarray}
\begin{split}
\nu &= \sqrt{1+m^2 r_5^2+p^2 r_5^2\lambda},\label{nu}=\sqrt{\nu_0^2+p^2 r_5^2\lambda}~,\\
X &=\frac{r_1}{r_5}p~,
\end{split}
\end{eqnarray}
with 
\begin{eqnarray}\label{nu0}
\nu_0\equiv \nu(\lambda=0)=\sqrt{1+m^2 r_5^2}~.
\end{eqnarray}
 The asymptotic expansion (\ie\ $z\to 0$) of the above solution \eqref{sol} is given by
\begin{eqnarray}\label{psiasmt}
\psi(z)\sim z^{\Delta_+}\left( \left(\frac{X}{2}\right)^{\Delta_+}\Gamma(-\nu)+O(z^2)\right)+z^{\Delta_-}\left( \left(\frac{X}{2}\right)^{\Delta_-}\Gamma(\nu)+O(z^2)\right)~,
\end{eqnarray}
where
\begin{eqnarray}\label{deltapm}
\Delta_{\pm}&=& 1\pm \nu~.
\end{eqnarray}
For normalizable solutions we must have 
\begin{eqnarray}\label{norsol}
\lim_{z\to 0}\psi(z)=\lim_{z\to 0}\psi'(z)=0~.
\end{eqnarray}
Requiring the exponent of $z$ to be real, one  recovers the modified Breitenlohner-Freedman (BF)  bound on  mass squared as
\begin{eqnarray}\label{BFbound}
\nu^2\geq0 \ \implies\ m^2+p^2\lambda \geq -\frac{1}{r_5^2}~.
\end{eqnarray}
That the BF bound depends on the square of the energy-momentum vector is a signature of the non-locality of the spacetime theory. When the BF bound is satisfied, $\Delta_+\geq 1$ and the mode $z^{\Delta_+}$ is always normalizable.

From the above analysis one can draw the following conclusions:
\begin{eqnarray}
\begin{split}
\text{for marginal operator }\mathcal{O}: \ \ \Delta_+=2 & \ \ \implies m^2=-p^2\lambda~,\\
\text{for relevant operator }\mathcal{O}: \ \ \Delta_+<2 & \ \ \implies m^2<-p^2\lambda~, \\
\text{for irrelevant operator }\mathcal{O}: \ \ \Delta_+>2 & \ \ \implies m^2>-p^2\lambda~.
\end{split}
\end{eqnarray}
Using standard holographic dictionary, one can identify $\Delta = \Delta_+$ as the dimension of the operator $\mathcal{O}(x)$ of the spacetime theory sourced by some current $J(x)$. Using the standard holographic dictionary, one can identify $\Delta_+=2h_p$ given in  \eqref{hhp} with $t=l_s^2\lambda$. As mentioned before, $\lambda$ can be chosen to have any positive value. For $\lambda=1$, $t=l_s^2$ as read off by comparing the spectrum \eqref{ttbarspect} with that of single trace $T\bar{T}$ deformed  string theory in $AdS_3$ in the untwisted sector.  For future convenience, let us also write
\begin{eqnarray}\label{dict}
2h_p=1+\nu~.
\end{eqnarray}

\subsubsection{Euclidean correlation function}
 Next, we aim to calculate the two-point function of operators in the spacetime theory. 
Let us start by introducing certain notations: the normalizable mode of the massive scalar $\Psi$ as $z\to 0$ is denoted by $\Psi_n$ and the non-normalizable mode is denoted by $\Psi_{nn}$.
The Euclidean partition function of the boundary theory with source $J(x)$ is given by
\begin{eqnarray}
Z_{CFT}[J(x)]=\left\langle e^{\int d^2x J(x)\mathcal{O}(x)}\right\rangle_E~.
\end{eqnarray}
According to gauge gravity duality
\begin{eqnarray}
Z_{CFT}[J(x)]=Z_{gravity}\left[\Psi_{nn}(x,z)=J(x)z^{\Delta_-}\right]~.
\end{eqnarray}
In the decoupling limit $g_s\to 0$, the leading contribution of the bulk partition function is given by its stationary point i.e. the classical solution
\begin{eqnarray}
Z_{gravity}=e^{S_E[\Psi_c]} ~,
\end{eqnarray}
where $\Psi_c$ is the classical solution to equation of motion that satisfies the appropriate boundary condition. The Euclidean classical action $S_E[\Psi_c]$ at the stationary point  is given by
\begin{eqnarray}
S_E[\Psi_c]&=& \int d^3 X \sqrt{-g}e^{-2\Phi}\left( R+4g^{\mu\nu}\partial_\mu\Phi\partial_\nu\Phi-\frac{1}{12}H^2-4\Lambda-\frac{1}{2}\left(g^{\mu\nu}\partial_\mu\Psi_c\partial_\nu\Psi_c+m^2\Psi_c^2\right)\right)~.\nonumber\\
\end{eqnarray}
Thus in the decoupling limit one can write
\begin{eqnarray}\label{holo}
\log Z_{CFT}[J]=S_E[\Psi_c]~.
\end{eqnarray}
The LHS of the equality in \eqref{holo} has the usual UV divergence of the a field theory. This corresponds to the volume divergence in the $\mathcal{M}_3$ on the RHS \footnote{The same volume divergence is also present in $AdS_3$.}. Thus one needs to regularize the volume divergence by imposing a cutoff at $z\to\epsilon$ for small $\epsilon$ and add a counter term:
\begin{eqnarray}
S^{(R)}_E[\Psi_c]=S_E[\Psi_c]\big{|}_{z=\epsilon}+S_{ct}[\Psi_c(\epsilon)]~,
\end{eqnarray}
where $S_E^{(R)}[\Psi_c]$ is the renormalized action and $S_{ct}[\Psi_c(\epsilon)]$ is the counter term added to make the renormalized action finite. Thus the connected $n$-point connected correlation function of local operators of the boundary theory is given by \footnote{The suffix $c$ in the LHS of \eqref{corr} denotes connected correlation function.}
\begin{eqnarray}\label{corr}
\langle \mathcal{O}(x_1)\cdots \mathcal{O}(x_n)\rangle_c=\frac{\delta^n\log Z^{(R)}_{CFT}}{\delta J(x_1)\cdots \delta J(x_n)}\Bigg{|}_{J=0}=\frac{\delta^n S^{(R)}_{E}[\Psi_c]}{\delta J(x_1)\cdots \delta J(x_n)}\Bigg{|}_{J=0}.
\end{eqnarray}

Next, we will focus on computing $S^{(R)}_E$ in the limit $\lim_{z\to 0}z^{-\Delta_-}\Psi_{c}(x,z)\to J(x)$. Let us denote the Lagrangian for the matter field $\Psi$ as
\begin{eqnarray}\label{matterlag}
S_m[\Psi]=-\frac{1}{2}\int d^2xdz \sqrt{-g}e^{-2\Phi}\left( g^{\mu\nu}\partial_\mu\Psi\partial_\nu\Psi+m^2\Psi^2\right).
\end{eqnarray}
The conjugate momenta, $\Pi$, is given by
\begin{eqnarray}\label{canmom}
\Pi=\frac{\delta S_m}{\delta\partial_z\Psi}=-e^{-2\Phi}\sqrt{g}g^{zz}\partial_z\Psi~.
\end{eqnarray}
The action $S_m$ \eqref{matterlag} evaluated at the classical solution $\Psi_c$ is given by
\begin{eqnarray}
\begin{split}
S_m&= \frac{1}{2}\int d^2x dz \Psi_c\underbrace{\left(\partial_\mu(\sqrt{g}e^{-2\Phi}g^{\mu\nu}\partial_\nu\Psi_c)-e^{-2\Phi}\sqrt{g}m^2\Psi^2_c\right)}_{=0\text{ by e.o.m}}-\frac{1}{2}\int d^2x dz \partial_\mu(\sqrt{g}e^{-2\Phi}g^{\mu\nu}\partial_\nu\Psi_c\Psi_c) \\
&= -\frac{1}{2}\int d^2 x(\sqrt{g}e^{-2\Phi}g^{\mu\nu}\partial_\nu\Psi_c(x,z))\Psi_c(x,z)\Big{|}_0^\infty\\
&= \frac{1}{2}\int d^2 x\Pi_c(x,z)\Psi_c(x,z)\Big{|}_0^\infty\\
&=\frac{1}{2} \int \frac{d^2 p}{(2\pi)^2}\Psi_c(p,z)\Pi_c(-p,z)\Big{|}_0^\infty~.
\end{split}
\end{eqnarray}
The classical solution $\Psi_c$ and its derivative $\partial_z\Psi_c$, both go to zero at $z\to\infty$.  This can be checked explicitly from \eqref{sol}. Moreover one can check that $\Pi_c$ in \eqref{canmom} goes to 0 as $z\to\infty$. Near $z=0$, the scalar field $\Psi_c$ behaves as 
\begin{eqnarray}\label{asypsi}
\Psi_c\to A(p)z^{\Delta_-}+B(p)z^{\Delta_+}~,
\end{eqnarray}
where $A$ and $B$ are smooth functions of the boundary momentum $p$ \footnote{From the standard holographic dictionary, $A(p)=J(p)$ and $B(p)$ is related to the one point function of the operator $\mathcal{O}$ \eqref{oneptfun}.}. Then \eqref{canmom} gives the behavior of $\Pi_c$ near the boundary as
\begin{eqnarray}
\Pi_c\to r_5\left(A(p)\Delta_- z^{\Delta_--2}+B(p)\Delta_+z^{\Delta_+-2}\right) .
\end{eqnarray}
Since $\Delta_+\geq1$, the term $\Psi_c\Pi_c$ in the action is divergent. Thus one needs, to introduce a counter term to cancel this divergence. The renormalized action $S^{(R)}_m[\Psi_c]$ takes the form 
\begin{eqnarray}
S^{(R)}_m[\Psi_c]=-\frac{1}{2}\int d^2 x \Psi_c\Pi_c\Big{|}_{z=\epsilon}+S_{ct}[\Psi_c(x,\epsilon)]~,
\end{eqnarray}
where the counter term takes the form
\begin{eqnarray}
S_{ct}[\Psi_c(x,\epsilon)]=\frac{1}{2}\int \frac{d^2p}{(2\pi)^2}F(p^2)\Psi_c(p,z)\Psi_c(-p,z) ~,
\end{eqnarray}
where $F$ is some analytic function of $p^2$ chosen such that it cancels the divergence in the action when $\epsilon$ is taken to $0$. To fix $F$, one needs to consider terms of sub-leading order in \eqref{asypsi}. Let us consider  a basis of functions $\Psi_+$ and $\Psi_-$ such that 
\begin{eqnarray}
\Psi_c(k,z)=A(p)\Psi_-(z)+B(p)\Psi_+(z)~.
\end{eqnarray}
The basis $\Psi_+$ and $\Psi_-$ must have the property that for small $z$
\begin{eqnarray}
\begin{split}
\Psi_+ &= z^{\Delta_+}(1+a_1p^2z^2+\cdots)~,\\
\Psi_- &= z^{\Delta_-}(1+b_1p^2z^2+\cdots)~,
\end{split}
\end{eqnarray}
where $a_i$ and $b_i$ are constant coefficients which will not be important for our computation. Similarly one can consider a basis $\Pi_+$ and $\Pi_-$ such that 
\begin{eqnarray}
\Pi_c=r_5\left(A(p)\Pi_-(z)+B(p)\Pi_+(z)\right)~.
\end{eqnarray}
Thus the renormalized action takes the form:
\begin{eqnarray}\label{smr}
S^{(R)}_m[\Psi_c]=-\frac{r_5}{2}\int \frac{d^2p}{(2\pi)^2}\left(A^2\Pi_-\Psi_-+B^2\Pi_+\Psi_++AB(\Pi_-\Psi_++\Psi_-\Pi_+)\right)+S_{ct}[\Psi_c(k,\epsilon)]~, \nonumber \\
\end{eqnarray}
where we used the following notations: $A^2=A(p)A(-p), AB=A(p)B(-p), B^2=B(p)B(-p)$. In the above expression the first term is divergent, the second term vanishes and the third term is finite. Thus one can choose the counter term as
\begin{eqnarray}
\begin{split}\label{sct}
S_{ct}&=\frac{r_5}{2}\int_{z=\epsilon} \frac{d^2p}{(2\pi)^2}\frac{\Pi_-}{\Psi_-}\Psi^2\\
&=\frac{r_5}{2}\int_{z=\epsilon} \frac{d^2p}{(2\pi)^2} \left(A^2\Psi_-\Pi_-+2AB\Pi_-\Psi_++B^2\frac{\Pi_-}{\Psi_-}\Psi_+^2\right)~.
\end{split}
\end{eqnarray}
One can easily check that $F(p^2)=\frac{\Pi_-}{\Psi_-}\Big{|}_\epsilon$ is indeed and analytic function of $p^2$. Note that the first term in \eqref{sct} is divergent and cancels the divergent term in $S^{(R)}_m[\Psi_c]$ \eqref{smr}. The second term in \eqref{sct} is finite and the third term vanishes.  Substituting \eqref{sct} in \eqref{smr}, one obtains the renormalized action as
\begin{eqnarray}
S^{(R)}_m[\Psi_c]=\frac{r_5}{2}\int \frac{d^2p}{(2\pi)^2}2\nu A(-p)B(p)~.
\end{eqnarray}
Imposing the boundary condition that $A(p)=J(p)$ as $z\to 0$ and the regularity of the solution at $z\to\infty$ \footnote{To impose regularity of the solution, one needs to work with the exact solution \eqref{sol}.}, uniquely fixes $\chi(p)=B(p)/A(p)$.  Thus the renormalized action takes the following form
\begin{eqnarray}
S^{(R)}_m=\frac{r_5}{2}\int \frac{d^2p}{(2\pi)^2}2\nu \chi(p)J(p)J(-p)~.
\end{eqnarray}
Thus from \eqref{corr}, the one point function takes the form
\begin{eqnarray}\label{oneptfun}
\langle \mathcal{O}(p)\rangle=2\nu\chi(p)J(p)=2\nu B(p)~.
\end{eqnarray}
The two-point function is given by
\begin{eqnarray}\label{oocorr}
\langle\mathcal{O}(p)\mathcal{O}(-p)\rangle_c=\frac{\delta^2S_m^{(R)}}{\delta J(p)\delta J(-p)}=2\nu\chi(p)= \frac{2\nu B(p)}{A(p)}~.
\end{eqnarray}
To compute $A$ and $B$ we expand the Bessel function in \eqref{sol} near $z\to 0$ as
\begin{eqnarray}\label{kxz}
K_\nu(Xz)=\frac{\Gamma(\nu)}{2}\left(\frac{X}{2}\right)^{-\nu}(1+\cdots)+\frac{\Gamma(-\nu)}{2}\left(\frac{X}{2}\right)^{\nu}(1+\cdots)~.
\end{eqnarray}
Recall that $X=(r_1/r_5)p$. From \eqref{kxz} one can read off $A$ and $B$ as
\begin{eqnarray}\label{AB}
A(p)=\frac{\Gamma(\nu)}{2}\left(\frac{X}{2}\right)^{-\nu} \text{ and } \ \ B(p)= \frac{\Gamma(-\nu)}{2}\left(\frac{X}{2}\right)^{\nu}.
\end{eqnarray}
Substituting \eqref{AB} in \eqref{oocorr}, one can write the correlation function as 
\begin{eqnarray}\label{g2pt}
\langle\mathcal{O}(p)\mathcal{O}(-p)\rangle_c= 2\nu \frac{\Gamma(-\nu)}{\Gamma(\nu)}\left(\frac{X}{2}\right)^{2\nu}=2\nu \frac{\Gamma(-\nu)}{\Gamma(\nu)}\left(\frac{r_1}{r_5}\right)^{2\nu}\left(\frac{p^2}{4}\right)^{\nu},
\end{eqnarray}
where $\nu$ is given by \eqref{nu}.  The two-point function \eqref{g2pt} matches exactly with the one calculated from the worldsheet approach \eqref{doo2ptmom} upon using the dictionary \eqref{dict}. Note that setting $\lambda=0$, exactly reproduces the two-point function of local operators of CFT$_2$ dual to string theory in $AdS_3$. As has been discussed at the end of section \ref{sec5.1}, the part $p^{2\nu}$  in the two-point function \eqref{g2pt} is the most interesting part of the two-point function because by appropriate normalization of the operators of the undeformed theory (\ie\ string theory in $AdS_3$), one can absorb the pre-factor in the redefinition of the operators.

\section{Discussion} \label{sec6}

In this paper, we derived the sigma model background of the coset $\frac{SL(2,\mathbb{R})_k\times U(1)}{U(1)}$ with the gauge currents \eqref{gaugecurr} and showed that this describes the full near horizon theory of the NS5 branes in a system of $k$ NS5 branes, $p$ F1 strings and $n$ units of momentum along the compact direction on which wraps the F1 strings both at zero and finite temperatures.  To be more precise, we established an equivalence among the followings: (a) string worldsheet sigma model on \ $\frac{SL(2,\mathbb{R})_k\times U(1)}{U(1)}\times SU(2)_k\times U(1)^4$ with the gauge currents \eqref{gaugecurr}, (b) the full near horizon theory of the NS5 branes with $k$ NS5 branes wrapping $T^4\times S^1$, $p$ F1 strings wrapping $S^1$ and $n$ units of momentum along the $S^1$ and (c) single trace $T\bar{T}$ deformation of string theory in $AdS_3\times S^3\times T^4$.  As a check to the above proposed equivalence, we computed the spectrum of the spacetime theory using the BRST quantization of the coset description and showed that it exactly matches with the one computed in the case of single trace $T\bar{T}$  deformation of string theory in $AdS_3$. Secondly, we computed the two-point correlation function of local operators of the spacetime theory using worldsheet techniques and reproduced the same two-point function using supergravity approach. The two-point function exhibits interesting analytic structure that are absent in a local quantum field theory. 

The coset construction has the following advantages. In the coset construction one can systematically construct the physical vertex operators and hence the physical Hilbert space. Calculation of the partition function by summing over the characters and imposing the BRST constraints is much simpler in the coset description. One can carry out a more thorough modular analysis of the torus partition function in the coset description and investigate the different phases of the theory.  It is also easy to extend the worldsheet techniques of section \ref{sec5.1} to compute the three point functions of the spacetime theory. It would be interesting to understand the analytic properties of the three point function of the spacetime theory. The coset construction may even help us understanding the asymptotically flat regime of the NS5 brane geometry. The sigma model background \eqref{bkg2} can also be obtained starting from BTZ black hole and performing a sequence of T-duality-shift-T-duality (TsT) in the background geometry \cite{Apolo:2019zai,Sfondrini:2019smd}.\footnote{We thank the JHEP referee for raising this point.} Indeed it would be interesting to understand a direct relation (if there exist any) between a coset model of the form $G/H$ with TsT, but unfortunately to the best of our knowledge, this issue has not been understood as of now. It is not clear if the coset model that we studied in this paper is, in some sense, very special, that can also be realized as performing  a particular TsT  transformation \cite{Apolo:2019zai,Sfondrini:2019smd} on $AdS_3/$BTZ. The spacetime theory, dual to string theory in the undeformed background \ie\ $AdS_3$ (or more precisely massless BTZ), in our setup, is in the R vacuum. It is not clear how one can study the deformation starting from the NS vacuum. One particular issue that arises working with the NS vacuum (\eg\ in the coordinates used in \cite{Maldacena:2000hw}) is that $J^-\bar{J}^-$ deformation (also known as single trace $T\bar{T}$ deformation) gives rise to a complex geometry. Similar issues arise in the coset approach as well if the the $SL(2,\mathbb{R})$ spacetime theory is in its NS vacuum. TsT on the other hand works perfectly fine in the case of both vacua \cite{Apolo:2019zai}. It would be interesting to understand the issue, raised above, in more detail.

%One intriguing fact is that the spectrum of long strings on $\frac{SL(2,\mathbb{R})_k\times U(1)}{U(1)}$ is same as those obtained in the case of single trace $T\bar{T}$ deformed string theory in $AdS_3$ regardless of which $U(1)_L\times U(1)_R$ is gauged. This is possibly related to the fact that $\frac{SL(2,\mathbb{R})_k\times U(1)}{U(1)}$ is a $J^-\bar{J}^-$ deformation of the worldsheet theory in $AdS_3$. It would be interesting to understand a more precise relation between similar coset descriptions involving $SL(2,\mathbb{R})$ and current-anti-current deformation of the worldsheet theory in $AdS_3$.

The coset description studied in this paper corresponds to that particular sign of the coupling of the single trace $T\bar{T}$ deformation of  string theory in $AdS_3$ that corresponds to a unitary spectrum. It would be nice to understand the coset description for the other sign of the coupling. It has been shown in \cite{Chakraborty:2020swe}, that for the other sign of the coupling the dual geometry corresponds to the full near horizon theory of the NS5 branes in a system of $k$ NS5 branes with $p$ negative F1 strings and $n$ units of momentum modes along the compact direction on which negative F1 strings wrap. The background geometry contains a naked singularity with closed timelike curves in a region beyond a certain radial distance. It has been argued in \cite{Chakraborty:2020swe}, that for the black hole thermodynamics to make sense, all one needs to care about are the states with real energies. It would be interesting to understand this phenomenon from the coset description point of view. Discarding the states with complex energies from the spectrum would imply giving up modular invariance. The coset description may come in handy in understanding the violation of the modular invariance.

Recently \cite{Eberhardt:2018ouy}, it has been proposed that string theory in $AdS_3\times S^3\times T^4$ with one (\ie\ $k=1$) unit of NS-NS flux is holographically dual to $(T^4)^N/S_N$. The usual RNS formalism of string theory in $AdS_3\times S^3\times T^4$ brakes down at $k=1$. This is due to the fact that the central charge of the $SU(2)$ WZW model on $S^3$ becomes negative at $k=1$. The way to tackle this issue is to employ the hybrid formalism where string theory on $AdS_3\times S^3$ is replaced by super WZW model on $PSU(1,1|2)$ at $k=1$. The single trace $T\bar{T}$ deformation of string theory is $AdS_3\times S^3\times T^4$ for $k\geq2$ heavily relies on the RNS formalism. Naive analytic continuation to the case $k=1$ does not always work. It would be interesting to study the single trace $T\bar{T}$ deformation of string theory in $AdS_3\times S^3\times T^4$ at $k=1$ using the coset approach. It is worth investigating the coset $\frac{PSU(1,1|2)_1\times U(1)}{U(1)}$ in hybrid formalism with the $SL(2,\mathbb{R})$ level $k=1$ and calculate the spectrum. If the spectrum in the winding one sector (or equivalently the untwisted sector of $(T^4)^N/S_N$)  agrees with the spectrum of a double trace $T\bar{T}$ deformed CFT$_2$ \cite{Smirnov:2016lqw,Cavaglia:2016oda} then one can consider $\frac{PSU(1,1|2)_1\times U(1)}{U(1)}\times U(1)^4$ as the single trace $T\bar{T}$ deformation of string theory in $AdS_3\times S^3\times T^4$ at $k=1$. It may also lead us to the theory of one single NS5 brane.

In the past few years, there has been studies of a Lorentz symmetry breaking solvable deformation of string theory in $AdS_3$ that goes in the name of single trace $J\bar{T}$ deformation \cite{Chakraborty:2018vja,Apolo:2018qpq,Apolo:2019yfj}. It would be nice to have a have an anomaly free coset description of such a deformation. One novelty of this coset description is that it allows us to study the deformed theory at finite temperature as well. It would also be nice to have a coset description that corresponds to  a deformation by a general linear combination of single trace $T\bar{T},J\bar{T},T\bar{J}$ of string theory in $AdS_3\times S^1$ \cite{Araujo:2018rho,Chakraborty:2019mdf,Chakraborty:2020xyz,Chakraborty:2020cgo,Chakraborty:2020udr}. Although the string background obtained after a deformation  by a general linear combination of single trace $T\bar{T},J\bar{T},T\bar{J}$ is known at zero temperature \cite{Chakraborty:2019mdf}, the geometry at finite temperature is not  known yet. The coset description will give the background at finite temperature. Another possible way to generate the finite temperature background is to start with the sigma model on $\frac{SL(2,\mathbb{R})_k\times U(1)}{U(1)}\times U(1)$ and then perform the sequence of TsT's listed in section 4 of  \cite{Chakraborty:2020udr}.

\appendix
\section{Parametrization of $SL(2,\mathbb{R})$}\label{appA}

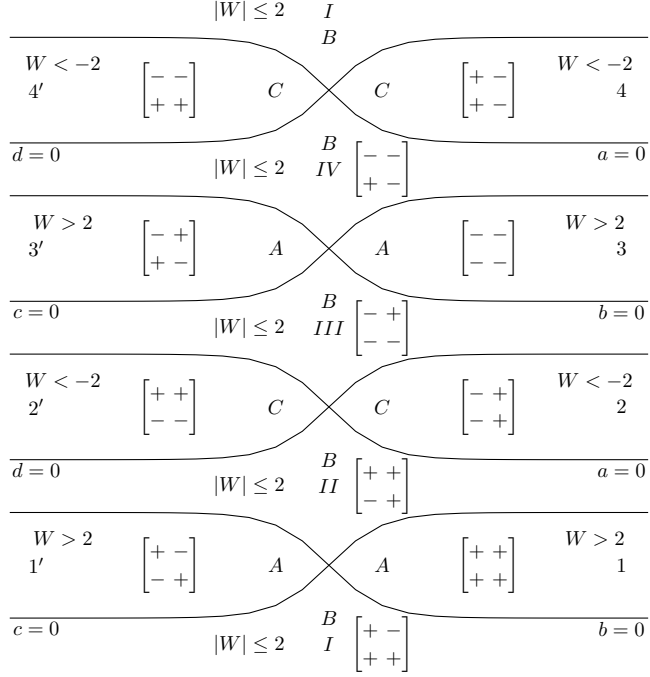
\begin{figure}
 \begin{center}
\begin{tikzpicture}[scale=.7, transform shape]

\draw [thick] [black,thin,domain=-6:6] plot ({\x}, {tanh(\x)});
\draw [thick] [black,thin,domain=-6:6] plot ({\x}, {tanh(-\x)});
\draw [thick] [black,thin,domain=-6:6] plot ({\x}, {tanh(\x)+3});
\draw [thick] [black,thin,domain=-6:6] plot ({\x}, {tanh(-\x)+3});
\draw [thick] [black,thin,domain=-6:6] plot ({\x}, {tanh(\x)+6});
\draw [thick] [black,thin,domain=-6:6] plot ({\x}, {tanh(-\x)+6});
\draw [thick] [black,thin,domain=-6:6] plot ({\x}, {tanh(\x)-3});
\draw [thick] [black,thin,domain=-6:6] plot ({\x}, {tanh(-\x)-3});
\draw (0,1.5) node {$III$};
\draw (0,4.5) node {$IV$};
\draw (0,7.5) node {$I$};
\draw (0,-1.5) node {$II$};
\draw (0,-4.5) node {$I$};

\draw (0,2) node {$B$};
\draw (0,5) node {$B$};
\draw (0,7) node {$B$};
\draw (0,-1) node {$B$};
\draw (0,-4) node {$B$};

\draw (1,0) node {$C$};
\draw (1,3) node {$A$};
\draw (1,6) node {$C$};
\draw (1,-3) node {$A$};

\draw (-1,0) node {$C$};
\draw (-1,3) node {$A$};
\draw (-1,6) node {$C$};
\draw (-1,-3) node {$A$};

\draw (-1.5,1.5) node {$|W|\leq 2$};
\draw (-1.5,4.5) node {$|W|\leq 2$};
\draw (-1.5,7.5) node {$|W|\leq 2$};
\draw (-1.5,-1.5) node {$|W|\leq 2$};
\draw (-1.5,-4.5) node {$|W|\leq 2$};

\draw (5,3.5) node {$W> 2$};
\draw (5,6.5) node {$W<-2$};
\draw (5,0.5) node {$W<-2$};
\draw (5,-2.5) node {$W>2$};

\draw (-5,3.5) node {$W> 2$};
\draw (-5,6.5) node {$W<-2$};
\draw (-5,0.5) node {$W<-2$};
\draw (-5,-2.5) node {$W>2$};

\draw (-5.5,-1.2) node {$d=0$};
\draw (-5.5,1.8) node {$c=0$};
\draw (-5.5,4.8) node {$d=0$};
\draw (-5.5,-4.2) node {$c=0$};

\draw (5.5,-1.2) node {$a=0$};
\draw (5.5,1.8) node {$b=0$};
\draw (5.5,4.8) node {$a=0$};
\draw (5.5,-4.2) node {$b=0$};

\draw (5.5,0) node {$2$};
\draw (5.5,3) node {$3$};
\draw (5.5,6) node {$4$};
\draw (5.5,-3) node {$1$};

\draw (-5.5,0) node {$2'$};
\draw (-5.5,3) node {$3'$};
\draw (-5.5,6) node {$4'$};
\draw (-5.5,-3) node {$1'$};

\draw (3,0) node {$\begin{bmatrix}
-&+\\
-&+
\end{bmatrix}$};
\draw (3,3) node {$\begin{bmatrix}
-&-\\
-&-
\end{bmatrix}$};
\draw (3,6) node {$\begin{bmatrix}
+&-\\
+&-
\end{bmatrix}$};
\draw (3,-3) node {$\begin{bmatrix}
+&+\\
+&+
\end{bmatrix}$};

\draw (-3,0) node {$\begin{bmatrix}
+&+\\
-&-
\end{bmatrix}$};
\draw (-3,3) node {$\begin{bmatrix}
-&+\\
+&-
\end{bmatrix}$};
\draw (-3,6) node {$\begin{bmatrix}
-&-\\
+&+
\end{bmatrix}$};
\draw (-3,-3) node {$\begin{bmatrix}
+&-\\
-&+
\end{bmatrix}$};

\draw (1,-1.5) node {$\begin{bmatrix}
+&+\\
-&+
\end{bmatrix}$};
\draw (1,1.5) node {$\begin{bmatrix}
-&+\\
-&-
\end{bmatrix}$};
\draw (1,4.5) node {$\begin{bmatrix}
-&-\\
+&-
\end{bmatrix}$};
\draw (1,-4.5) node {$\begin{bmatrix}
+&-\\
+&+
\end{bmatrix}$};
\end{tikzpicture}
 \end{center}
\caption{Two dimensional slice of $SL(2,\mathbb{R})$.}
  \label{12regions}
\end{figure}

In this appendix, we give a brief review of parametrization of $SL(2,\mathbb{R})$ \cite{Elitzur:2002rt}. An element $g\in SL(2,\mathbb{R})$ is given by
\begin{eqnarray}
g=\begin{pmatrix}
a&& b\\
 c&& d
\end{pmatrix}~,
\end{eqnarray}
where $a,b,c,d\in\mathbb{R}$ and $ad-bc=1$. The group manifold $SL(2,\mathbb{R})$ splits into the following twelve regions (see figure \ref{12regions}):
\begin{itemize}
\item{Region A: $ad>0,bc>0$. The entries of the matrix $g$ has the following signs: $$\left\{\begin{pmatrix}
+&&+\\+&&+
\end{pmatrix},\begin{pmatrix}
+&&-\\-&&+
\end{pmatrix},\begin{pmatrix}
-&&-\\-&&-
\end{pmatrix},\begin{pmatrix}
-&&+\\+&&-
\end{pmatrix}\right\}~.$$
The regions $1,1',3,3'$ in figure \ref{12regions} with signs of the entries of the matrix $g$ given above respectively, are of type (A).}
\item{Region B: $ad>0,bc<0$. The entries of the matrix $g$ has the following signs: $$\left\{\begin{pmatrix}
+&&-\\+&&+
\end{pmatrix},\begin{pmatrix}
+&&+\\-&&+
\end{pmatrix},\begin{pmatrix}
-&&+\\-&&-
\end{pmatrix},\begin{pmatrix}
-&&-\\+&&-
\end{pmatrix}\right\}~.$$
The regions $I,II,III,IV$ in figure \ref{12regions} with signs of the entries of the matrix $g$ given above respectively, are of type (B)}
\item{Region C: $ad<0,bc<0$. The entries of the matrix $g$ has the following signs: $$\left\{\begin{pmatrix}
-&&+\\-&&+
\end{pmatrix},\begin{pmatrix}
+&&+\\-&&-
\end{pmatrix},\begin{pmatrix}
+&&-\\+&&-
\end{pmatrix},\begin{pmatrix}
-&&-\\+&&+
\end{pmatrix}\right\}~.$$
The regions $2,2',4,4'$ in figure \ref{12regions} with signs of the entries of the matrix $g$ given above respectively, are of type (C).}
\end{itemize} 
Let us define the quantity 
\begin{eqnarray}
W=\tr(\sigma_3 g\sigma_3 g^{-1})=2(2ad-1)=2(2bc+1)~,
\end{eqnarray}
which is invariant under action of the gauge group (In our case the gauge group is the non-compact $U(1)$ subgroup of $SL(2,\mathbb{R})$ that we wish to gauge.). The quantity $W>2$ in region (A), $|W|<2$ in region (B) and $W<-2$ in region (C).  Figure \ref{12regions} demonstrates the $12$ distinct regions. The boundary of $AdS_3$ corresponds to large $a,b,c,d$, \ie\ $|W|\to\infty$. From figure \ref{12regions} we see that this corresponds to region (A) and (C). The region (B) on the other hand doesn't contain the boundary of $AdS_3$. 

Given that $SL(2,\mathbb{R})$ can be divided into 12 regions (see figure \ref{12regions}), an element $g\in SL(2,\mathbb{R})$ in Lorentzian signature can be parametrized as
\begin{eqnarray}\label{para}
g=e^{\alpha\sigma_3}(-1)^{\epsilon_1}(i\sigma_2)^{\epsilon_2}g_\delta(\theta)e^{\beta\sigma_3}~,
\end{eqnarray}
where $\sigma_i$ are that Pauli matrices given by
\begin{eqnarray}
\sigma_1=\begin{pmatrix}
0&&1\\ 1&&0
\end{pmatrix},\ \ \ \ \ \sigma_2=\begin{pmatrix}
0&&-i\\ i&&0
\end{pmatrix},\ \ \ \ \ \sigma_3=\begin{pmatrix}
1&&0\\ 0&&-1
\end{pmatrix}~,
\end{eqnarray}
and
\begin{eqnarray}
\epsilon_{1,2}=\{0,1\}, \ \ \  \delta=1,1',I ~.
\end{eqnarray}
For regions $1,1',I$, $\epsilon_1=\epsilon_2=0$ and 
\begin{eqnarray}
&&g_I=\begin{pmatrix}
\cos\theta && -\sin\theta\\
\sin\theta && \cos\theta
\end{pmatrix};   \ \ \ \ 0\leq\theta\leq\pi/2~,\\
&&g_1=g_{1'}^{-1}= \begin{pmatrix}
\cosh\theta && \sinh\theta\\
\sinh\theta && \cosh\theta
\end{pmatrix};   \ \ \ \ 0\leq\theta<\infty~.
\end{eqnarray}
It is easy to check that the entries of $g$ in \eqref{para} are reals and determinant of $g$ is unity.
For the other nine regions $\epsilon_{1,2}$ takes different values. 

The parametrization in region (A), (B), (C) can be easily related to each other by analytic continuation in $\theta$. Let us denote $\theta=\theta_B$ in region (B). The parametrization in region (A) is obtained by $\theta_B\to i\theta_A$ and the parametrization of region (C) is obtained by  $\theta_B\to i\theta_C+\pi/2$.

\section*{Acknowledgements} 
The author would like to thank A. Gadde, A. Giveon, A. Hashimoto, D. Kutasov, S. Minwalla and S. Roy for many helpful discussions and valuable comments on the manuscripts. The author is supported by the Infosys Endowment for the study of the Quantum Structure of Spacetime.

\newpage

%\bibliography{ref}\bibliographystyle{JHEP}

\begin{thebibliography}{10}

\bibitem{Smirnov:2016lqw}
F.~Smirnov and A.~Zamolodchikov, \emph{{On space of integrable quantum field
  theories}},
  \href{https://doi.org/10.1016/j.nuclphysb.2016.12.014}{\emph{Nucl. Phys. B}
  {\bfseries 915} (2017) 363}
  [\href{https://arxiv.org/abs/1608.05499}{{\ttfamily 1608.05499}}].

\bibitem{Cavaglia:2016oda}
A.~Cavagli\`a, S.~Negro, I.~M. Sz\'ecs\'enyi and R.~Tateo, \emph{{$T
  \bar{T}$-deformed 2D Quantum Field Theories}},
  \href{https://doi.org/10.1007/JHEP10(2016)112}{\emph{JHEP} {\bfseries 10}
  (2016) 112} [\href{https://arxiv.org/abs/1608.05534}{{\ttfamily
  1608.05534}}].

\bibitem{Giveon:2017nie}
A.~Giveon, N.~Itzhaki and D.~Kutasov, \emph{{$ \mathrm{T}\overline{\mathrm{T}}
  $ and LST}}, \href{https://doi.org/10.1007/JHEP07(2017)122}{\emph{JHEP}
  {\bfseries 07} (2017) 122}
  [\href{https://arxiv.org/abs/1701.05576}{{\ttfamily 1701.05576}}].

\bibitem{Giveon:2017myj}
A.~Giveon, N.~Itzhaki and D.~Kutasov, \emph{{A solvable irrelevant deformation
  of AdS$_{3}$/CFT$_{2}$}},
  \href{https://doi.org/10.1007/JHEP12(2017)155}{\emph{JHEP} {\bfseries 12}
  (2017) 155} [\href{https://arxiv.org/abs/1707.05800}{{\ttfamily
  1707.05800}}].

\bibitem{Asrat:2017tzd}
M.~Asrat, A.~Giveon, N.~Itzhaki and D.~Kutasov, \emph{{Holography Beyond AdS}},
  \href{https://doi.org/10.1016/j.nuclphysb.2018.05.005}{\emph{Nucl. Phys. B}
  {\bfseries 932} (2018) 241}
  [\href{https://arxiv.org/abs/1711.02690}{{\ttfamily 1711.02690}}].

\bibitem{Giribet:2017imm}
G.~Giribet, \emph{{$T\bar{T}$-deformations, AdS/CFT and correlation
  functions}}, \href{https://doi.org/10.1007/JHEP02(2018)114}{\emph{JHEP}
  {\bfseries 02} (2018) 114}
  [\href{https://arxiv.org/abs/1711.02716}{{\ttfamily 1711.02716}}].

\bibitem{Chakraborty:2018kpr}
S.~Chakraborty, A.~Giveon, N.~Itzhaki and D.~Kutasov, \emph{{Entanglement
  beyond AdS}},
  \href{https://doi.org/10.1016/j.nuclphysb.2018.08.011}{\emph{Nucl. Phys. B}
  {\bfseries 935} (2018) 290}
  [\href{https://arxiv.org/abs/1805.06286}{{\ttfamily 1805.06286}}].

\bibitem{Chakraborty:2018aji}
S.~Chakraborty, \emph{{Wilson loop in a $T\bar{T}$ like deformed
  $\rm{CFT}_2$}},
  \href{https://doi.org/10.1016/j.nuclphysb.2018.12.003}{\emph{Nucl. Phys. B}
  {\bfseries 938} (2019) 605}
  [\href{https://arxiv.org/abs/1809.01915}{{\ttfamily 1809.01915}}].

\bibitem{Apolo:2019zai}
L.~Apolo, S.~Detournay and W.~Song, \emph{{TsT, $T\bar{T}$ and black strings}},
  \href{https://doi.org/10.1007/JHEP06(2020)109}{\emph{JHEP} {\bfseries 06}
  (2020) 109} [\href{https://arxiv.org/abs/1911.12359}{{\ttfamily
  1911.12359}}].

\bibitem{Chakraborty:2020swe}
S.~Chakraborty, A.~Giveon and D.~Kutasov, \emph{{$T\bar T$, Black Holes and
  Negative Strings}},
  \href{https://doi.org/10.1007/JHEP09(2020)057}{\emph{JHEP} {\bfseries 09}
  (2020) 057} [\href{https://arxiv.org/abs/2006.13249}{{\ttfamily
  2006.13249}}].

\bibitem{Chakraborty:2020fpt}
S.~Chakraborty, G.~Katoch and S.~R. Roy, \emph{{Holographic Complexity of LST
  and Single Trace $T\bar{T}$}},
  \href{https://arxiv.org/abs/2012.11644}{{\ttfamily 2012.11644}}.

\bibitem{Apolo:2018qpq}
L.~Apolo and W.~Song, \emph{{Strings on warped AdS$_{3}$ via $
  \mathrm{T}\bar{\mathrm{J}} $ deformations}},
  \href{https://doi.org/10.1007/JHEP10(2018)165}{\emph{JHEP} {\bfseries 10}
  (2018) 165} [\href{https://arxiv.org/abs/1806.10127}{{\ttfamily
  1806.10127}}].

\bibitem{Chakraborty:2018vja}
S.~Chakraborty, A.~Giveon and D.~Kutasov, \emph{{$ J\overline{T} $ deformed
  CFT$_{2}$ and string theory}},
  \href{https://doi.org/10.1007/JHEP10(2018)057}{\emph{JHEP} {\bfseries 10}
  (2018) 057} [\href{https://arxiv.org/abs/1806.09667}{{\ttfamily
  1806.09667}}].

\bibitem{Araujo:2018rho}
T.~Araujo, E.~Colg\'ain, Y.~Sakatani, M.~Sheikh-Jabbari and H.~Yavartanoo,
  \emph{{Holographic integration of $T \bar{T}$ \textbackslash{}\& $J \bar{T}$
  via $O(d,d)$}}, \href{https://doi.org/10.1007/JHEP03(2019)168}{\emph{JHEP}
  {\bfseries 03} (2019) 168}
  [\href{https://arxiv.org/abs/1811.03050}{{\ttfamily 1811.03050}}].

\bibitem{Chakraborty:2019mdf}
S.~Chakraborty, A.~Giveon and D.~Kutasov, \emph{{$T\bar{T}$, $J\bar{T}$,
  $T\bar{J}$ and String Theory}},
  \href{https://doi.org/10.1088/1751-8121/ab3710}{\emph{J. Phys. A} {\bfseries
  52} (2019) 384003} [\href{https://arxiv.org/abs/1905.00051}{{\ttfamily
  1905.00051}}].

\bibitem{Apolo:2019yfj}
L.~Apolo and W.~Song, \emph{{Heating up holography for single-trace $J\bar{T}$
  deformations}}, \href{https://doi.org/10.1007/JHEP01(2020)141}{\emph{JHEP}
  {\bfseries 01} (2020) 141}
  [\href{https://arxiv.org/abs/1907.03745}{{\ttfamily 1907.03745}}].

\bibitem{Chakraborty:2020cgo}
S.~Chakraborty, A.~Giveon and D.~Kutasov, \emph{{Strings in irrelevant
  deformations of AdS$_{3}$/CFT$_{2}$}},
  \href{https://doi.org/10.1007/JHEP11(2020)057}{\emph{JHEP} {\bfseries 11}
  (2020) 057} [\href{https://arxiv.org/abs/2009.03929}{{\ttfamily
  2009.03929}}].

\bibitem{Chakraborty:2020udr}
S.~Chakraborty and A.~Hashimoto, \emph{{Entanglement Entropy for $T \bar T$, $J
  \bar T$, $T \bar J$ deformed holographic CFT}},
  \href{https://arxiv.org/abs/2010.15759}{{\ttfamily 2010.15759}}.

\bibitem{Carlip:2005zn}
S.~Carlip, \emph{{Conformal field theory, (2+1)-dimensional gravity, and the
  BTZ black hole}},
  \href{https://doi.org/10.1088/0264-9381/22/12/R01}{\emph{Class. Quant. Grav.}
  {\bfseries 22} (2005) R85}
  [\href{https://arxiv.org/abs/gr-qc/0503022}{{\ttfamily gr-qc/0503022}}].

\bibitem{Dijkgraaf:1991ba}
R.~Dijkgraaf, H.~L. Verlinde and E.~P. Verlinde, \emph{{String propagation in a
  black hole geometry}},
  \href{https://doi.org/10.1016/0550-3213(92)90237-6}{\emph{Nucl. Phys. B}
  {\bfseries 371} (1992) 269}.

\bibitem{Giveon:2003ge}
A.~Giveon, E.~Rabinovici and A.~Sever, \emph{{Beyond the singularity of the 2-D
  charged black hole}},
  \href{https://doi.org/10.1088/1126-6708/2003/07/055}{\emph{JHEP} {\bfseries
  07} (2003) 055} [\href{https://arxiv.org/abs/hep-th/0305140}{{\ttfamily
  hep-th/0305140}}].

\bibitem{Giveon:2005mi}
A.~Giveon, D.~Kutasov, E.~Rabinovici and A.~Sever, \emph{{Phases of quantum
  gravity in AdS(3) and linear dilaton backgrounds}},
  \href{https://doi.org/10.1016/j.nuclphysb.2005.04.015}{\emph{Nucl. Phys. B}
  {\bfseries 719} (2005) 3}
  [\href{https://arxiv.org/abs/hep-th/0503121}{{\ttfamily hep-th/0503121}}].

\bibitem{Giveon:2006pr}
A.~Giveon and D.~Kutasov, \emph{{Fundamental strings and black holes}},
  \href{https://doi.org/10.1088/1126-6708/2007/01/071}{\emph{JHEP} {\bfseries
  01} (2007) 071} [\href{https://arxiv.org/abs/hep-th/0611062}{{\ttfamily
  hep-th/0611062}}].

\bibitem{Giveon:2019fgr}
A.~Giveon, \emph{{Comments on $T\bar T$, $J\bar{T}$ and String Theory}},
  \href{https://arxiv.org/abs/1903.06883}{{\ttfamily 1903.06883}}.

\bibitem{Baggio:2018gct}
M.~Baggio and A.~Sfondrini, \emph{{Strings on NS-NS Backgrounds as Integrable
  Deformations}}, \href{https://doi.org/10.1103/PhysRevD.98.021902}{\emph{Phys.
  Rev. D} {\bfseries 98} (2018) 021902}
  [\href{https://arxiv.org/abs/1804.01998}{{\ttfamily 1804.01998}}].

\bibitem{Dei:2018mfl}
A.~Dei and A.~Sfondrini, \emph{{Integrable spin chain for stringy
  Wess-Zumino-Witten models}},
  \href{https://doi.org/10.1007/JHEP07(2018)109}{\emph{JHEP} {\bfseries 07}
  (2018) 109} [\href{https://arxiv.org/abs/1806.00422}{{\ttfamily
  1806.00422}}].

\bibitem{Giveon:1998ns}
A.~Giveon, D.~Kutasov and N.~Seiberg, \emph{{Comments on string theory on
  AdS(3)}}, \href{https://doi.org/10.4310/ATMP.1998.v2.n4.a3}{\emph{Adv. Theor.
  Math. Phys.} {\bfseries 2} (1998) 733}
  [\href{https://arxiv.org/abs/hep-th/9806194}{{\ttfamily hep-th/9806194}}].

\bibitem{Kutasov:1999xu}
D.~Kutasov and N.~Seiberg, \emph{{More comments on string theory on AdS(3)}},
  \href{https://doi.org/10.1088/1126-6708/1999/04/008}{\emph{JHEP} {\bfseries
  04} (1999) 008} [\href{https://arxiv.org/abs/hep-th/9903219}{{\ttfamily
  hep-th/9903219}}].

\bibitem{Giveon:2001up}
A.~Giveon and D.~Kutasov, \emph{{Notes on AdS(3)}},
  \href{https://doi.org/10.1016/S0550-3213(01)00573-9}{\emph{Nucl. Phys. B}
  {\bfseries 621} (2002) 303}
  [\href{https://arxiv.org/abs/hep-th/0106004}{{\ttfamily hep-th/0106004}}].

\bibitem{Maldacena:2000hw}
J.~M. Maldacena and H.~Ooguri, \emph{{Strings in AdS(3) and SL(2,R) WZW model
  1.: The Spectrum}}, \href{https://doi.org/10.1063/1.1377273}{\emph{J. Math.
  Phys.} {\bfseries 42} (2001) 2929}
  [\href{https://arxiv.org/abs/hep-th/0001053}{{\ttfamily hep-th/0001053}}].

\bibitem{Maldacena:2000kv}
J.~M. Maldacena, H.~Ooguri and J.~Son, \emph{{Strings in AdS(3) and the SL(2,R)
  WZW model. Part 2. Euclidean black hole}},
  \href{https://doi.org/10.1063/1.1377039}{\emph{J. Math. Phys.} {\bfseries 42}
  (2001) 2961} [\href{https://arxiv.org/abs/hep-th/0005183}{{\ttfamily
  hep-th/0005183}}].

\bibitem{Maldacena:2001km}
J.~M. Maldacena and H.~Ooguri, \emph{{Strings in AdS(3) and the SL(2,R) WZW
  model. Part 3. Correlation functions}},
  \href{https://doi.org/10.1103/PhysRevD.65.106006}{\emph{Phys. Rev. D}
  {\bfseries 65} (2002) 106006}
  [\href{https://arxiv.org/abs/hep-th/0111180}{{\ttfamily hep-th/0111180}}].

\bibitem{Seiberg:1999xz}
N.~Seiberg and E.~Witten, \emph{{The D1 / D5 system and singular CFT}},
  \href{https://doi.org/10.1088/1126-6708/1999/04/017}{\emph{JHEP} {\bfseries
  04} (1999) 017} [\href{https://arxiv.org/abs/hep-th/9903224}{{\ttfamily
  hep-th/9903224}}].

\bibitem{Eberhardt:2018ouy}
L.~Eberhardt, M.~R. Gaberdiel and R.~Gopakumar, \emph{{The Worldsheet Dual of
  the Symmetric Product CFT}},
  \href{https://doi.org/10.1007/JHEP04(2019)103}{\emph{JHEP} {\bfseries 04}
  (2019) 103} [\href{https://arxiv.org/abs/1812.01007}{{\ttfamily
  1812.01007}}].

\bibitem{Forste:1994wp}
S.~Forste, \emph{{A Truly marginal deformation of SL(2, R) in a null
  direction}}, \href{https://doi.org/10.1016/0370-2693(94)91340-4}{\emph{Phys.
  Lett. B} {\bfseries 338} (1994) 36}
  [\href{https://arxiv.org/abs/hep-th/9407198}{{\ttfamily hep-th/9407198}}].

\bibitem{Israel:2003ry}
D.~Israel, C.~Kounnas and M.~P. Petropoulos, \emph{{Superstrings on NS5
  backgrounds, deformed AdS(3) and holography}},
  \href{https://doi.org/10.1088/1126-6708/2003/10/028}{\emph{JHEP} {\bfseries
  10} (2003) 028} [\href{https://arxiv.org/abs/hep-th/0306053}{{\ttfamily
  hep-th/0306053}}].

\bibitem{Goykhman:2013oja}
M.~Goykhman and A.~Parnachev, \emph{{Stringy holography at finite density}},
  \href{https://doi.org/10.1016/j.nuclphysb.2013.05.011}{\emph{Nucl. Phys. B}
  {\bfseries 874} (2013) 115}
  [\href{https://arxiv.org/abs/1304.4496}{{\ttfamily 1304.4496}}].

\bibitem{Karabali:1989dk}
D.~Karabali and H.~J. Schnitzer, \emph{{BRST Quantization of the Gauged WZW
  Action and Coset Conformal Field Theories}},
  \href{https://doi.org/10.1016/0550-3213(90)90075-O}{\emph{Nucl. Phys. B}
  {\bfseries 329} (1990) 649}.

\bibitem{Tseytlin:1993my}
A.~A. Tseytlin, \emph{{Conformal sigma models corresponding to gauged
  Wess-Zumino-Witten theories}},
  \href{https://doi.org/10.1016/0550-3213(94)90461-8}{\emph{Nucl. Phys. B}
  {\bfseries 411} (1994) 509}
  [\href{https://arxiv.org/abs/hep-th/9302083}{{\ttfamily hep-th/9302083}}].

\bibitem{Bastianelli:1990ey}
F.~Bastianelli, \emph{{BRST symmetry from a change of variables and the gauged
  WZNW models}},
  \href{https://doi.org/10.1016/0550-3213(91)90273-Z}{\emph{Nucl. Phys. B}
  {\bfseries 361} (1991) 555}.

\bibitem{Elitzur:2002rt}
S.~Elitzur, A.~Giveon, D.~Kutasov and E.~Rabinovici, \emph{{From big bang to
  big crunch and beyond}},
  \href{https://doi.org/10.1088/1126-6708/2002/06/017}{\emph{JHEP} {\bfseries
  06} (2002) 017} [\href{https://arxiv.org/abs/hep-th/0204189}{{\ttfamily
  hep-th/0204189}}].

\bibitem{Parsons:2009si}
J.~Parsons and S.~F. Ross, \emph{{Strings in extremal BTZ black holes}},
  \href{https://doi.org/10.1088/1126-6708/2009/04/134}{\emph{JHEP} {\bfseries
  04} (2009) 134} [\href{https://arxiv.org/abs/0901.3044}{{\ttfamily
  0901.3044}}].

\bibitem{Argurio:2000tb}
R.~Argurio, A.~Giveon and A.~Shomer, \emph{{Superstrings on AdS(3) and
  symmetric products}},
  \href{https://doi.org/10.1088/1126-6708/2000/12/003}{\emph{JHEP} {\bfseries
  12} (2000) 003} [\href{https://arxiv.org/abs/hep-th/0009242}{{\ttfamily
  hep-th/0009242}}].

\bibitem{Sfondrini:2019smd}
A.~Sfondrini and S.~J. van Tongeren, \emph{{$T\bar{T}$ deformations as $TsT$
  transformations}},
  \href{https://doi.org/10.1103/PhysRevD.101.066022}{\emph{Phys. Rev. D}
  {\bfseries 101} (2020) 066022}
  [\href{https://arxiv.org/abs/1908.09299}{{\ttfamily 1908.09299}}].

\bibitem{Chakraborty:2020xyz}
S.~Chakraborty and A.~Hashimoto, \emph{{Thermodynamics of $
  \mathrm{T}\overline{\mathrm{T}} $, $ \mathrm{J}\overline{\mathrm{T}} $, $
  \mathrm{T}\overline{\mathrm{J}} $ deformed conformal field theories}},
  \href{https://doi.org/10.1007/JHEP07(2020)188}{\emph{JHEP} {\bfseries 07}
  (2020) 188} [\href{https://arxiv.org/abs/2006.10271}{{\ttfamily
  2006.10271}}].

\end{thebibliography}

\providecommand{\href}[2]{#2}\begingroup\raggedright\endgroup

\end{document}